\documentclass[journal]{IEEEtran}

\usepackage{cite}
\usepackage{amsmath,amsthm,amssymb,amsfonts} 
\usepackage{algorithm}
\usepackage{algorithmic}
\usepackage{enumerate}
\usepackage{graphicx}
\usepackage{epstopdf}
\usepackage{url}
\usepackage{bm}
\usepackage{color}

\newtheorem{theorem}{Theorem}

\newtheorem{lemma}{Lemma}
\newtheorem{corollary}{Corollary}

\newtheorem{assumption}{Assumption}

\newcommand{\tabincell}[2]{
	\begin{tabular}{@{}#1@{}}
		
		#2
	\end{tabular}
}
\newcommand{\sinc}{\operatorname{sinc}}

\IEEEoverridecommandlockouts

\begin{document}

\title{Physical Layer Security in Heterogeneous Networks with Jammer Selection and Full-Duplex Users}

\author{
	\IEEEauthorblockN{Weijun Tang, Suili Feng, Yuehua Ding, and Yuan Liu}
	\thanks{
				
		The authors are with School of Electronic and Information Engineering, South China University of Technology, Guangzhou 510641, China (e-mail: tang.weijun@mail.scut.edu.cn, \{fengsl, eeyhding, eeyliu\}@scut.edu.cn).
%

		Part of this paper was presented at the IEEE Global Communications Conference (GLOBECOM), Washington, DC, USA, December, 2016 \cite{tang_jammer_2016}.}
}
\maketitle

\begin{abstract}
In this paper, we enhance physical layer security for downlink heterogeneous networks (HetNets) by using friendly jammers and full-duplex users.
The jammers are selected to transmit jamming signal if their interfering power on the scheduled users is below a threshold, meanwhile the scheduled users confound the eavesdroppers using artificial noise by full-duplexing.
Using the tools of stochastic geometry, we derive the expressions of connection probability and secrecy probability.
In particular, the locations of active jammers are modeled by a modified Poisson hole process (PHP).
Determining the jammer selection threshold is further investigated for connection probability maximization subject to the security constraints.
A greedy algorithm is proposed to efficiently solve this problem.
The accuracy of the theoretical analysis and the efficiency of the proposed algorithm are evaluated by numerical simulations.
\end{abstract}


\IEEEpeerreviewmaketitle

\section{Introduction}
With the data explosion in wireless traffic, the issue of privacy and security has become one of the most important concerns in wireless networks.
Due to the broadcast nature, wireless communication is very fragile to be eavesdropped.
Traditionally, security is usually focused by upper layers.
Recently, physical layer security has been regarded as a promising technology to complement and improve the security of wireless networks.
The basic idea of physical layer security is to exploit the physical characteristics of wireless channels to secure messages in an information-theoretical view.
In the pioneering work \cite{wyner_wire-tap_1975}, Wyner showed that if the main source-destination channel is better than the eavesdropping channel, the transmitted message can be perfectly secured at a non-zero rate.
To this end, relay, cooperative jamming and artificial noise are efficient approaches \cite{zhang_artificial_2016, zou_survey_2016, liu_power_2017}.

Recently, there has also been an increasing interest in heterogeneous cellular networks (HetNets) as a means to fulfill seamless wireless coverage and high network throughput in next generation wireless networks.
Due to the open system architecture and densification, information transmissions intended for authorized users are more vulnerable to eavesdroppers in HetNets.
Therefore, secure transmission is a significant concern when designing HetNets.

\subsection{Related Works}
Secure communications of one source-destination pair with multiple cooperating relays were investigated in \cite{dong_improving_2010}.
The authors proposed three cooperative schemes to maximize the achievable secrecy rate.
To protect the confidential message from non-colluding passive eavesdroppers, the authors in \cite{zhang_improving_2015} studied the secrecy performance for multi-input-multi-output (MISO) systems via transmitting artificial noise in the null space of legitimate channel.
Secure routing in multihop ad-hoc networks was studied in \cite{yao_secure_2016}.
The authors in \cite{wang_opportunistic_2016} proposed an opportunistic multiple jammer selection scheme for multi-eavesdropper scenario.
In \cite{liu_ergodic_2016}, the probability of secure connection and ergodic secrecy capacity of one source-destination pair with multiple cooperative jammers were studied, where a secrecy protected zone around the source node and a interferer-excluded zone around the destination node were designed.

The works \cite{dong_improving_2010, zhang_improving_2015, yao_secure_2016, wang_opportunistic_2016, liu_ergodic_2016} all considered one source-destination pair, and there was no interference between the legitimate nodes. 
Using the tools of stochastic geometry, the security performance of cellular networks considering the cell association and the information exchange between base stations (BSs) was evaluated in \cite{wang_physical_2013}.
This work was extended by \cite{geraci_physical_2014} for the downlink multiple-antenna cellular networks.
Assuming that the legitimate mobile users were potential eavesdroppers, the achievable secrecy rates were analyzed in \cite{geraci_physical_2014}.
The secure communication in uplink transmissions was investigated in \cite{deng_secure_2016}.
The closed-form expressions of the ergodic secrecy sum rates for a random user were studied.
The authors in \cite{deng_physical_2016} developed a tractable framework for the average secrecy rate in the three-tier wireless sensor networks consisting of remote sensors, access points and the sinks.
The authors in \cite{wang_physical_mmWave_2016} studied the artificial noise aided security performance in a millimeter wave network for both non-colluding and colluding eavesdroppers scenarios. 
The authors in \cite{wang_physical_2016} studied physical layer security in multi-tier HetNets, where an access-threshold-based secrecy user association policy was proposed.
A multiuser MISO network with full-duplex users was considered in \cite{akgun_exploiting_2017}, where the receivers transmit artificial noise for jamming the nearby eavesdroppers when received information signals.

\subsection{Motivations and Contributions} 
In this paper, we investigate a HetNet with friendly jammers and multiple non-colluding eavesdroppers, which has not been considered yet in the prior works.
For the sake of easy deployment, all nodes in the HetNet are equipped with single antenna, and the jammers are assumed to transmit artificial noise independently.
To alleviate the interference from jammers to users, we propose a jammer selection policy based on the received jamming power at the users.
That is, a jammer is selected to be active if its jamming strength on the scheduled users is below a threshold.
Unlike the scheme in \cite{wang_opportunistic_2016}, we consider the path loss effects in this paper since the large-scale fading is dominant in received signal strength.
Hence each scheduled user is associated with a guard zone around. 
To enhance security in the guard zones where the friendly jammers are silent, we assume that the users have the capability of full duplex so that they can confound the eavesdroppers by transmitting artificial noise when they receive the information from their serving BSs.
The main contributions of this paper are summarized as follow:
\begin{enumerate}
	\item A new physical layer security scheme in HetNets is designed, where jammer selection along with full-duplex users' artificial noise are used.
	Significant improvements of the system performance are obtained.
	\item Using stochastic geometry tools, we derive the tractable analyses of user connection probability and user secrecy probability.
	In particular, the locations of the active jammers are modeled by a modified Poisson hole process (PHP), where the baseline Poisson point process (PPP) of jammers are homogeneous and the locations of the holes (i.e. the scheduled users) are modeled by \emph{inhomogeneous} PPP.
	The lower bounds for the Laplace transform of interference from active jammers are obtained.
	\item The SINR thresholds for user connection and secrecy transmission and the jammer selection threshold are jointly optimized.
	The optimization problem is formulated as a nonlinear programming for user connection probability maximization subject to secrecy probability and secrecy rate constraints. 
	An efficient algorithm is proposed for solving this problem.
	\item Valuable insights are provided for practical designs.
	In particular, the user connection probability is independent of the associated tier, and the secrecy performance of the users in small cells is better than that in macro cells.
	The user connection probability is improved by increasing the jammer density and the jammer transmit power.
\end{enumerate}

The rest of this paper is organized as follows.
Section \ref{sec_system_model} presents the system model. 
Section \ref{sec_analytical_result} provides the analytical results. 
The optimization problem and the numerical results are shown in Section \ref{sec_maximization} and Section \ref{sec_numerical_results}, respectively.
Section \ref{sec_conclusion} concludes this paper.

\section{System Model}
\label{sec_system_model}
\begin{figure*}[t]
	\centering
	\includegraphics[width=160mm]{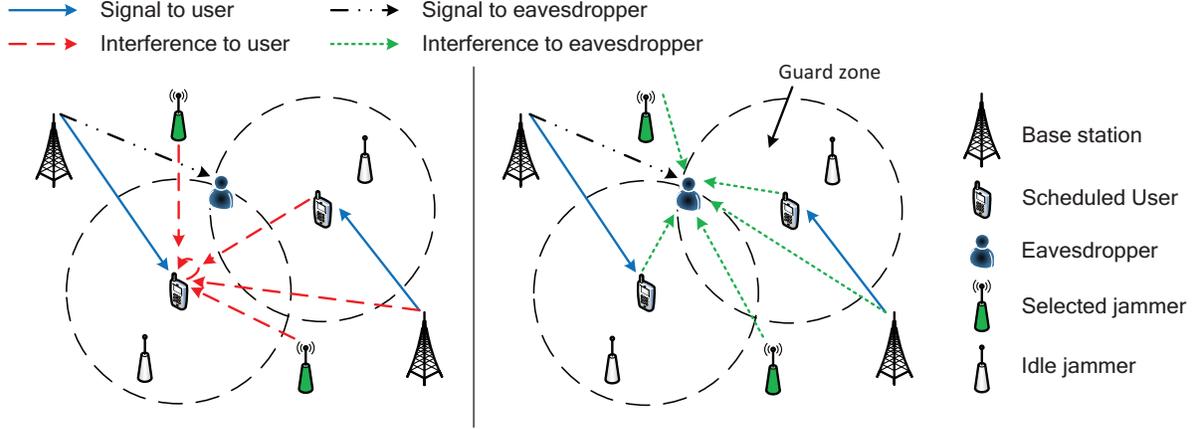}
	\vspace{-0.1cm}
	\caption{An instance of HetNet with the proposed jammer selection scheme and full-duplex users.}
	\label{fig_netowk}
	\vspace{-0.3cm}
\end{figure*}

We consider a downlink $K$-tier HetNet where the BSs in different tiers sharing the same frequency band.
Define $\mathcal{K} \triangleq \{1,2, \cdots K\}$.
The BS locations of tier $t$ are distributed as an independent homogeneous PPP $\Phi_t$ with density $\lambda_t$ in a two-dimensional plane $\mathbb{R}^2$.
As shown in Fig. \ref{fig_netowk}, there users and eavesdroppers coexist.
The users are authorized destinations which are full-duplex capable.
The eavesdroppers attempt to intercept the confidential information intended for the users.
Friendly jammers are deployed to ensure the secrecy transmission by sending artificial noise independently.
The users, the eavesdroppers and the jammers are spatially distributed according to independent homogeneous PPPs $\Phi_U$, $\Phi_E$ and $\Phi_J$ with densities $\lambda_U$, $\lambda_E$ and $\lambda_J$, respectively.
$\lambda_U$ is assumed to be much greater than $\{\lambda_t\}_{t \in \mathcal{K}}$ so that there are associated users in each BS.
Every BS in tier $t$ has the same transmit power $P_t$.
$P_J$ and $P_U$ are denoted as the transmit power of jammers and users, respectively.
All the nodes in this network are equipped with single antenna.

We consider both large-scale and small-scale fading for wireless channels.
The path loss model $l(d) = d^{-\alpha}$ is adopted for the large-scale fading, where $d$ is the propagation distance, and $\alpha > 2$ is the path loss exponent.
For the small-scale fading, we assume independent Rayleigh fading, and the channel gains follow the exponential distribution with unit power.
Note that shadow fading can be approximately modeled by the randomness of the node locations \cite{lin_optimizing_2015}.
Hence shadowing is omitted here for simplicity and tractability.

We consider an open access HetNet so that a user is associated with (the nearest BS in) tier $k$, if $k = \arg\max_{t \in \mathcal{K}} P_t D_t^{-\alpha}$, where $D_t$ is denoted as the distance between the user and its nearest BS in tier $t$.
The associated BS is called the serving BS. 
After the user association, each BS schedules its associated users according to time division multiple access (TDMA), i.e., scheduling one user in each time slot.
Therefore, there is no intra-cell interference (but inter-cell interference still exist due to full spectrum reuse).

\subsection{Jammer Selection and Full-Duplex User Jamming}
Jammers are placed into the network for securing the legitimate transmission.
However, since the locations and the channel state information (CSI) of passive eavesdroppers are unavailable, it is difficult to select the jammers according to their locations and CSI.
Moreover, the jammers simultaneously interfere with the users when confounding eavesdroppers, and thus degrade the user connection outage performance.
Hence, we propose a jammer selection scheme to protect users from over jamming.
A jammer is selected to be active if its interference to any scheduled user is below the jammer selection threshold $\tau$, i.e. $\Phi_J^s  = \{\mathbf{ j} | \mathbf{ j} \in \Phi_J, P_J D_{\mathbf{ j}, \mathbf{u}_i}^{-\alpha} < \tau, \forall \mathbf{u}_i \in \Phi_U^s\}$.
$\Phi_J^s$ denotes the point process of active jammers.
$D_{\mathbf{x}, \mathbf{y}}$ denotes the distance between node $\mathbf{x}$ and $\mathbf{y}$. 
$\Phi_U^s = \bigcup_{t \in \mathcal{K}} \Phi_{U, t}^s$, where $\Phi_{U, t}^s$ is the point process of scheduled users in tier $t$.
According to this scheme, there is a guard zone (also called hole in the rest of this paper) around each scheduled user, as shown in Fig. \ref{fig_netowk}. Denote the radius of each guard zone by $R_{\tau}$, we have
\begin{equation}
R_{\tau} = \left(\frac{P_J}{\tau}\right)^{1 / \alpha}.
\end{equation}
Hence $\Phi_J^s$ can be expressed as 
\begin{equation}
\Phi_J^s = \Phi_J \setminus \bigcup\limits_{\mathbf{y} \in \Phi_U^s} \mathbf{B}(\mathbf{y}, R_{\tau}), 
\end{equation}
where $\mathbf{B}(\mathbf{y}, R_{\tau}) \equiv \{\mathbf{z} \in \mathbb{R}^2 : \| \mathbf{z} - \mathbf{y} \| < R_{\tau}\}$ is a ball of radius $R_{\tau}$ centered at $\mathbf{y}$.
Therefore, $\Phi_J^s$ is a (modified) PHP which can be formally defined in terms of two independent point processes: $\Phi_J$ and $\Phi_U^s$.
$\Phi_J$ represents the baseline PPP from which the holes are carved out, and $\Phi_U^s$ represents the locations of holes.
More discussions about $\Phi_J^s$ are shown in the next section.

The proposed jammer selection scheme can restrain the interference to scheduled users from jammers.
However, it leaves jammer-excluded zones around the scheduled users, which can be exploited by nearby eavesdroppers.
Hence we assume that the users are full-duplex capable.
The scheduled users transmit artificial noise simultaneously when they are listening to their serving BSs.
Thanks to self-interference cancellation (SIC), the legitimate users can suffer low self-interference while the eavesdroppers are being confounded.
An instance of our scheme is shown in Fig. \ref{fig_netowk}. 
For ease of reading, the notations in this paper are summarized in Table \ref{table_notation}.

\begin{table}[t]
	\renewcommand{\arraystretch}{1.3}
	\caption{Notation Summary}
	\label{table_notation}
	\centering
	\begin{tabular}{|c|c|}
		\hline
		\bfseries Notation & \bfseries Description\\
		\hline
		\hline
		$\Phi_t, \Phi_U, \Phi_E, \Phi_J$ & \tabincell{c}{Point process of BSs in tier $t$ / users / \\ eavesdroppers / jammers}\\
		\hline
		$\Phi_{U}^{s}, \Phi_{U,t}^{s}, \Phi_{J}^{s}$ & \tabincell{c}{Point process of scheduled users / \\ scheduled users in tier $t$ / active jammers}\\
		\hline
		$\lambda_t, \lambda_U, \lambda_E, \lambda_J$ & \tabincell{c}{Density of BSs in tier $t$ / users / \\ eavesdroppers / jammers}\\
		\hline
		$\lambda_U^s, \lambda_{U,t}^s$ & \tabincell{c}{Density of scheduled users / \\ scheduled users in tier $t$}\\
		\hline
		$P_t, P_U, P_J$ & \tabincell{c}{Transmit power of BSs in tier $t$ / users / jammers}\\
		\hline
		$N_S; N_0$ & \tabincell{c}{residual self-interference; thermal noise}\\
		\hline
		$\tau; R_{\tau}$ & \tabincell{c}{Jammer selection threshold; \\the radius of guard zone}\\
		\hline
		$\alpha; \beta$ & \tabincell{c}{Path loss exponent; SIC capacity}\\
		\hline
		$D_t$ & \tabincell{c}{Distance from the typical user to \\its closest BS in tier $t$ }\\
		\hline
		$D_{\mathbf{i},\mathbf{j}}$ & \tabincell{c}{Distance between node $\mathbf{i}$ and node $\mathbf{j}$}\\
		\hline
		$h_{\mathbf{i},\mathbf{j}}$ & \tabincell{c}{Small scale fading between node $\mathbf{i}$ and node $\mathbf{j}$}\\
		\hline
		$\mathcal{A}_t$ & \tabincell{c}{Association probabilities of \\the typical user to tier $t$}\\
		\hline
		$I_B, I_U, I_J$ & \tabincell{c}{Cumulative interference on the typical \\user from BSs / scheduled users / active jammers}\\
		\hline
		$I'_B, I'_U, I'_J$ & \tabincell{c}{Cumulative interference on the eavesdropper \\from BSs / scheduled users / active jammers}\\
		\hline
		$\theta_c, \theta_s$ & \tabincell{c}{SINR threshold for user connection / \\ secrecy transmission}\\
		\hline
		$\mathcal{P}_c, \mathcal{P}_s$ & User connection / secrecy probability\\
		\hline
		$\mathcal{P}_T, \mathcal{R}_T$ & \tabincell{c}{Quality of service requirements of\\ secrecy probability / secrecy rate}\\
		\hline
		$\Delta$ & \tabincell{c}{Step size in Algorithm \ref{alg_thresholds_optimization}}\\
		\hline
	\end{tabular}
\end{table}

\section{Analytical Modeling and Results}
\label{sec_analytical_result}
In this section, we investigate the connection probability and the secrecy probability of a randomly located user.
Denote $\theta_c$ and $\theta_s$ as the SINR thresholds for user connection and secrecy transmission.
We assume that a user can decode the secret messages if its received SINR is greater than $\theta_c$.
On the other hand, an eavesdropper can not get any confidential information from the messages when its received SINR is less than $\theta_s$.

\subsection{User Connection Probability}
\label{subsection_user_connection_probability}
Without loss of generality, we consider a typical user located at the origin, which is denoted as $\mathbf{u}_0$ and its serving BS (also called the tagged BS) is $\mathbf{b}_0$.
The user connection probability is defined as
\begin{equation}
\label{eqn_user_connection_probability_definition}
\mathcal{P}_c(\theta_c, \tau) \triangleq \mathbb{P}\left(\text{SINR}_U \geq \theta_c \right),
\end{equation}
where $\text{SINR}_U$ is given by
\begin{align}
\text{SINR}_U = \frac{P_{\mathbf{b}_0} h_{\mathbf{b}_0, \mathbf{u}_0} D_{\mathbf{b}_0, \mathbf{u}_0}^{-\alpha}}{I_B + I_U + I_J + N_S +N_0}.
\end{align}
$P_{\mathbf{b}_0}$ is the transmit power of the tagged BS.
The small-scale fading between node $\mathbf{x}$ and $\mathbf{y}$ is denoted as $h_{\mathbf{x}, \mathbf{y}}$.
The cumulative interference from the BSs (except its serving BS) is given by $I_B = \sum_{t \in \mathcal{K}}\sum_{\mathbf{b}_i \in \Phi_t \setminus \mathbf{b}_0} P_t h_{\mathbf{b}_i, \mathbf{u}_0} D_{\mathbf{b}_i, \mathbf{u}_0}^{-\alpha}$.
Similarly, $I_U$ and $I_J$ are the cumulative interference from scheduled users and active jammers, respectively, which are given by $I_U = \sum_{t \in \mathcal{K}} \sum_{\mathbf{u}_i \in \Phi_{U, t}^s \setminus \mathbf{u}_0} P_U h_{\mathbf{u}_i, \mathbf{u}_0} D_{\mathbf{u}_i, \mathbf{u}_0}^{-\alpha}$ and $I_J = \sum_{\mathbf{z}_i \in \Phi_J^s} P_J h_{\mathbf{z}_i, \mathbf{u}_0} D_{\mathbf{z}_i, \mathbf{u}_0}^{-\alpha}$.
$N_S$ and $N_0$ represent the residual self-interference and the thermal noise, respectively.
In this paper, we assume a linear residual self-interference performance as $N_S = \beta P_U$, where $\beta$ is the SIC capacity.

To derive the theoretical analyses, we use some conclusions proposed in \cite{jo_heterogeneous_2012}, which are summarized as the following Lemma \ref{lemma_association_probability}, \ref{lemma_D_k_PDF} and \ref{lemma_Laplace_BS2user}.
\begin{lemma}
\label{lemma_association_probability}
The association probability that the serving BS of the typical user is in tier $k$ is given by
\begin{equation}
\label{eqn_association_probability}
\mathcal{A}_k = \frac{\lambda_k P_k^{2 / \alpha}}{\sum_{t \in \mathcal{K}} \lambda_t P_t^{2 / \alpha}}.
\end{equation} 
\end{lemma}

\begin{lemma}
\label{lemma_D_k_PDF}
Conditioned on $\mathbf{b}_0 \in \Phi_k$, the probability distribution function (PDF) of $D_{\mathbf{b}_0, \mathbf{u}_0}$ is given by 
\begin{equation}
\label{eqn_D_k_PDF}
f_{D_{\mathbf{b}_0, \mathbf{u}_0}} (r) = f_{D_k}(r) = \frac{2 \pi \lambda_k r}{\mathcal{A}_k} \exp\Big(- \frac{\pi r^2}{P_k^{2 / \alpha}} \Xi \Big),
\end{equation}
where $\Xi = \sum_{t \in \mathcal{K}} \lambda_t P_t^{2 / \alpha}$.
\end{lemma}

\begin{lemma}
\label{lemma_Laplace_BS2user}
Conditioned on $D_{\mathbf{b}_0, \mathbf{u}_0} = r$ and $\mathbf{b}_0 \in \Phi_k$, the Laplace transform of the interference from BSs to the typical user is 
\begin{align}
\label{eqn_Laplace_BS2user}
&\mathcal{L}_{I_B}(s) \nonumber\\
&= \prod_{t \in \mathcal{K}} \exp\Big( -2 \pi \lambda_t \frac{s P_t \Delta_{k, t}^{2 - \alpha}}{\alpha - 2} {}_2F_1(1, 1-\frac{2}{\alpha}; 2 - \frac{2}{\alpha}; -\frac{s P_t}{\Delta_{k, t}^{\alpha}})\Big),
\end{align}
where $\Delta_{k,t} = r (\frac{P_t}{P_k})^{1 / \alpha}$, and ${}_2F_1(a, b; c; d)$ is the Gauss hypergeometric function.
\end{lemma}


Due to the scheduling and association criteria, only one user per BS interferes with the typical user.
Therefore, $\{\Phi_{U, t}^s\}$ are not PPPs but Poisson-Voronoi perturbed lattice \cite{blaszczyszyn_clustering_2015}.
Moreover, there exists correlation between the typical user and the interfering nodes \cite{singh_joint_2015}.
To make it tractable, we characterize the scheduled users as inhomogeneous PPPs like in \cite{tang_hybrid_2016, singh_joint_2015}.

\begin{assumption}
	\label{assumption_intensity}
	Conditioned on $D_{\mathbf{b}_0, \mathbf{u}_0} = r$ and $\mathbf{b}_0 \in \Phi_k$, the point processes of scheduled users are assumed to be Poisson and independent with intensity measure function as
	\begin{align}
	\label{eqn_scheduled_user_intensity_tier_t}
	\lambda_{U, t}^s(r, y) &=  \lambda_t \Big(1 - \exp\Big(-\pi (r + y)^2 \lambda_t (\frac{P_t}{P_k})^{2 / \alpha}\Big)\Big),\\
	\label{eqn_scheduled_user_intensity}
	\lambda_{U}^s(r, y) &= \sum_{t \in \mathcal{K}} \lambda_{U, t}^s(r, y),
	\end{align}
	where $y$ is the distance to the typical user.
	Please refer to Appendix \ref{appendix_scheduled_user_intensity} for more details about these point processes. 
\end{assumption}

Using Assumption \ref{assumption_intensity}, we can have the Laplace transform of the interference from scheduled users to the typical user as shown in Lemma \ref{lemma_Laplace_user2user}.
\begin{lemma}
\label{lemma_Laplace_user2user}
Conditioned on $D_{\mathbf{b}_0, \mathbf{u}_0} = r$ and $\mathbf{b}_0 \in \Phi_k$, the Laplace transform of the interference from scheduled users to the typical user is 
\begin{align}
\label{eqn_Laplace_user2user}
&\mathcal{L}_{I_U}(s) \nonumber\\
&= \prod_{t \in \mathcal{K}} \exp \Big(-2\pi \int_{0}^{\infty} \lambda_{U, t}^s (r, y) (1 - \mathcal{L}_h(s P_U y^{-\alpha})) y dy \Big),
\end{align}
where $\mathcal{L}_h(s) = \frac{1}{1 + s}$ is the Laplace transform of $h \sim \exp(1)$.
\end{lemma}
\begin{IEEEproof}
\begin{align}
&\mathcal{L}_{I_{U}}(s) = \mathbb{E}\Big[\exp(-s(\sum_{t \in \mathcal{K}} \sum_{\mathbf{u}_i \in \Phi_{U, t}^s \setminus \mathbf{u}_0} P_U h_{\mathbf{u}_i, \mathbf{u}_0} D_{\mathbf{u}_i, \mathbf{u}_0}^{-\alpha})) \Big] \nonumber \\
&\overset{(a)}{=} \prod_{t \in \mathcal{K}} \mathbb{E}_{\Phi_{U, t}^s} \Big[\prod_{\mathbf{u}_i \in \Phi_{U, t}^s \setminus \mathbf{u}_0} \mathcal{L}_h (s P_U D_{\mathbf{u}_i, \mathbf{u}_0}^{-\alpha})\Big] \nonumber \\
&\overset{(b)}{=} \prod_{t \in \mathcal{K}} \exp \Big(-2\pi \int_{0}^{\infty} \lambda_{U, t}^s (r, y) (1 - \mathcal{L}_h(s P_U y^{-\alpha})) y dy \Big),
\end{align}
where $(a)$ follows that the small-scale fading $h$ and the point processes $\{\Phi_{U, t}^s\}$ are mutually independent.
$(b)$ is obtained by using the probability generating functional (PGFL) of PPP \cite{stoyan_stochastic_2013}.
\end{IEEEproof}

We now introduce our approach to characterize the Laplace transform of interference from active jammers.
We model the locations of jammers by a homogeneous PPP $\Phi_J$ of density $\lambda_J$.
The guard zones with radius $R_{\tau}$ are carved out according to the jammer selection scheme.
The locations of the centers of these holes are the scheduled users $\Phi_U^s$.
As discussed above, $\Phi_U^s$ is modeled as an inhomogeneous PPP.
Therefore $\Phi_J^s$ is not a standard PHP which can be defined in terms of two independent \emph{homogeneous} PPPs \cite{yazdanshenasan_poisson_2016}.
In this paper, we call the type of point process as $\Phi_J^s$ \emph{modified} PHP.
A modified PHP can be defined in terms of two independent PPPs: the baseline PPP and the locations of the holes.
The holes are carved out from the baseline PPP (e.g. $\Phi_J$) which is homogeneous, while the locations of the holes (e.g. $\Phi_{U}^s$) are \emph{inhomogeneous}.
On the same lines as discussed in \cite{yazdanshenasan_poisson_2016}, characterizing the interference experienced by the typical user in a (modified) PHP precisely is complicated.
Hence, we consider only two holes: (i) the one around the typical user, and (ii) the one that is closest to the typical users, and ignore the other holes.
This is reasonable because the influence of the carved-holes which are far from the typical user is limited due to the path loss.
This approach captures the local neighborhood of the typical user accurately and provides an upper bound of the interference from active jammers.
Using this approach, we derive the Laplace transform of interference from active jammers as shown in Lemma \ref{lemma_Laplace_jammer2user}.
\begin{lemma}
\label{lemma_Laplace_jammer2user}
Conditioned on $D_{\mathbf{b}_0, \mathbf{u}_0} = r$ and $\mathbf{b}_0 \in \Phi_k$, the Laplace transform of the interference from active jammers to the typical user is 
\begin{align}
\label{eqn_Laplace_jammer2user}
&\mathcal{L}_{I_J}(s) \nonumber\\
&= \exp\Big(- 2 \pi \lambda_J \frac{s P_J R_{\tau}^{2 - \alpha}}{\alpha - 2} {}_2F_1(1, 1 - \frac{2}{\alpha}; 2 - \frac{2}{\alpha}; -\frac{s P_J}{R_{\tau}^{\alpha}})\Big) \nonumber\\ &\quad \times \int_{0}^{\infty} H(v, s)  f(v) dv,
\end{align}
where
\begin{align}
\label{eqn_H_function}
H(v, s) &= \exp\Big( \int_{\max\{v - R_{\tau}, R_{\tau}\}}^{v + R_{\tau}} \frac{2 y\lambda_J'(y)}{1 + \frac{y^{\alpha}}{s P_J}} dy \Big),\\
f(v) &= 2 \pi \lambda_U^s(r, v) v \exp(-2\pi \int_{0}^{v} \lambda_U^s(r, y) y dy),\\
\lambda_J'(y) &= \lambda_J\arccos(\frac{y^2 + v^2 - R_{\tau}^2}{2 y v}).
\end{align}
\end{lemma}
\begin{IEEEproof}
Please see Appendix \ref{appendix_laplace_jammer2user}.
\end{IEEEproof}

\begin{theorem}
\label{theorem_connection_probability}
The user connection probability of a typical user is given by
\begin{align}
\label{eqn_connection_probability}
\mathcal{P}_c(\theta_c, \tau) &= \sum_{k \in \mathcal{K}} \mathcal{A}_k \int_{0}^{\infty} \exp(- (N_S + N_0) \zeta_{k}(r)) \mathcal{L}_{I_B}(\zeta_{k}(r))\nonumber \\
& \qquad \qquad \qquad \times \mathcal{L}_{I_U}(\zeta_{k}(r)) \mathcal{L}_{I_J}(\zeta_{k}(r)) f_{D_k}(r) dr ,
\end{align}
where $\zeta_{k}(r) = \frac{\theta_c r^{\alpha}}{P_k}$. 
$\mathcal{L}_{I_B}(s)$, $\mathcal{L}_{I_U}(s)$, $\mathcal{L}_{I_J}(s)$ and $f_{D_k}(r)$ are \eqref{eqn_Laplace_BS2user}, \eqref{eqn_Laplace_user2user}, \eqref{eqn_Laplace_jammer2user} and \eqref{eqn_D_k_PDF}, respectively.  
\end{theorem}
\begin{IEEEproof}
Note that $I_U$ and $I_J$ are correlated because the locations of the scheduled users are also the centers of the guard zones.
However, the baseline point process $\Phi_J$ and the locations of the scheduled users are mutually independent.
Besides, we only consider the two closest holes of the active jammers.
Thus the correlation of $I_U$ and $I_J$ is weak, which is neglected here for the sake of tractability.
Then we have
\begin{align}
&\mathbb{P}(\text{SINR}_U \geq \theta_c | \mathbf{b}_0 \in \Phi_k, D_{\mathbf{b}_0, \mathbf{u}_0} = r) \nonumber \\
&= \mathbb{P}(\frac{P_k h_{\mathbf{b}_0, \mathbf{u}_0} r^{-\alpha}}{I_B + I_U + I_J + N_S + N_0} \geq \theta_c) \nonumber\\
&\overset{(a)}{=} \exp(- (N_S + N_0)\zeta_{k}(r)) \nonumber \\
& \qquad \times \mathcal{L}_{I_B}(\zeta_{k}(r))\mathcal{L}_{I_U}(\zeta_{k}(r)) \mathcal{L}_{I_J}(\zeta_{k}(r)),
\end{align}
where $(a)$ follows due to the assumption that the small-scale fading is i.i.d. exponential distribution with unit power, and $I_B$, $I_U$ and $I_J$ are mutually independent.
Then it is easy to get \eqref{eqn_connection_probability} by applying
\begin{align}
&\mathcal{P}_c(\theta_c, \tau) \nonumber\\
&= \sum_{k \in \mathcal{K}} \mathcal{A}_k \mathbb{E}_{D_{\mathbf{b}_0, \mathbf{u}_0}}[\mathbb{P}(\text{SINR}_U \geq \theta_c | \mathbf{b}_0 \in \Phi_k, D_{\mathbf{b}_0, \mathbf{u}_0} = r)].
\end{align}
\end{IEEEproof}

One can see that, the expression of the user connection probability in Theorem \ref{theorem_connection_probability} is very complicated.
To obtain some interesting insights, we adopt some approaches to simplify the result.

\subsubsection{Simplified $\mathcal{L}_{I_U}$}
\label{subsubsec_simplified_Laplace_user2user}
To provide a simplified expression of the Laplace transform $\mathcal{L}_{I_U}$, we model the locations of the interfering users as homogeneous PPPs.
This assumption has been widely adopted in the literatures \cite{elsawy_stochastic_2014, lee_hybrid_2015, liu_tractable_2016}.
Compared with the inhomogeneous PPP assumption, this approach is more tractable but less accurate.
\begin{lemma}
	\label{lemma_Laplace_user2user_simplified}
	If the locations of interfering users are modeled with homogeneous PPPs, conditioned on  $\mathbf{b}_0 \in \Phi_k$, the Laplace transform of the interference from the scheduled users to the typical user is 
	\begin{equation}
	\mathcal{L}_{I_U}(s) = \exp \Big(- \frac{\pi P_U^{2 / \alpha}\sum_{t \in \mathcal{K}} \lambda_t}{\sinc(2 / \alpha)} s^{2 / \alpha} \Big).
	\end{equation}
\end{lemma}
\begin{IEEEproof}
	In each time slot, there is only one user scheduled in each BS.
	Thus we have the intensity of the interfering users in tier $t$ (i.e. $\Phi_{U,t}^s$) as $\lambda_{U,t}^s = \lambda_t$.
	The Laplace transform $\mathcal{L}_{I_U}$ is
	\begin{align}
	\mathcal{L}_{I_U}(s) & = \exp \Big(-2 \pi \sum_{t \in \mathcal{K}} \lambda_{t} \int_{0}^{\infty} \frac{y}{1 + \frac{y^{\alpha}}{s P_U}} dy \Big) \nonumber\\
	& \overset{(a)}{=} \exp\Big(\frac{2}{\alpha} \pi (s P_U)^{2 / \alpha} \mathcal{B}(\frac{2}{\alpha}, 1 - \frac{2}{\alpha})\sum_{t \in \mathcal{K}} \lambda_t\Big),
	\end{align}
	where $(a)$ follows by replacing $x = y^{\alpha}$ and $\int_{0}^{\infty} \frac{x^{\mu - 1} dx}{(1 + \beta x)^{\nu}} = \beta^{-\mu}\mathcal{B}(\mu, \nu - \mu)$ \cite[Eq. 3.194.3]{gradshtein_table_2007}.
	After some algebra derivation and using $\mathcal{B}(x, 1 - x) = \frac{\pi}{\sin(\pi x)}$, we have the result.
\end{IEEEproof}

\subsubsection{Simplified $\mathcal{L}_{I_J}$}
To provide a simplified expression of the Laplace transform $\mathcal{L}_{I_J}$, we consider the case without jammer selection (i.e. $\tau = \infty$).
Then we have Lemma \ref{lemma_Laplace_jammer2user_simplified}.
\begin{lemma}
	\label{lemma_Laplace_jammer2user_simplified}
	If there is no jammer selection, conditioned on $\mathbf{b}_0 \in \Phi_k$, the Laplace transform of the interference from the jammers to the typical user is
	\begin{equation}
	\label{eqn_Laplace_jammer2user_simplified}
	\mathcal{L}_{I_J}(s) = \exp \Big(-\frac{\pi \lambda_J P_J^{2 / \alpha}}{\sinc(2 / \alpha)} s^{2 / \alpha} \Big).
	\end{equation}
\end{lemma}
\begin{IEEEproof}
	\begin{align}
	\mathcal{L}_{I_J}(s) &\overset{(a)}{=} \mathbb{E}_{\Phi_J} \Big[\prod_{\mathbf{z}_i \in \Phi_J} \mathcal{L}_h(s P_J D_{\mathbf{z}_i, \mathbf{u}_0}^{-\alpha}) \Big] \nonumber \\
	& = \exp\Big(-2 \pi \lambda_J \int_{0}^{\infty} \frac{y}{1 + \frac{y^{\alpha}}{s P_J}} dy\Big),
	\end{align}
	where $(a)$ follows by that there is no jammer selection.	
	Then \eqref{eqn_Laplace_jammer2user_simplified} is obtained following a similar process in the proof of Lemma \ref{lemma_Laplace_user2user_simplified}. 
\end{IEEEproof}

\begin{corollary}
	\label{corollary_user_connection_probability_simplified}
	When $N_S = N_0 = 0$ and the locations of the interfering users are approximated with homogeneous PPPs, the user connection probability without jammer selection is 
	\begin{align}
	\label{eqn_user_connection_probability_simplified}
	\tilde{\mathcal{P}}_c(\theta_c) = &\pi \Xi \int_{0}^{\infty} \exp\Big(-\frac{2 \pi v \theta_c \Xi}{\alpha - 2}{}_2F_1(1, 1 - \frac{2}{\alpha}; 2 - \frac{2}{\alpha}; \theta_c)\Big) \nonumber\\
	&\times \exp\Big(- \frac{\pi  v(\theta_c P_U)^{2 / \alpha} \sum_{t \in \mathcal{K}} \lambda_t}{\sinc(2 / \alpha)}\Big) \nonumber \\
	&\times \exp\Big(-\frac{\pi v (\theta_c P_J)^{2 / \alpha} \lambda_J}{\sinc(2 / \alpha)}\Big) \nonumber\\
	&\times \exp(-\pi v \Xi ) dv,
	\end{align}
	where $\Xi = \sum_{t \in \mathcal{K}} \lambda_t P_t^{2 / \alpha}$.
\end{corollary}
\begin{IEEEproof}
	Using the results in Lemma \ref{lemma_D_k_PDF}, \ref{lemma_Laplace_BS2user}, \ref{lemma_Laplace_user2user_simplified} and \ref{lemma_Laplace_jammer2user_simplified} and let $v = \frac{r^2}{P_k}$, we can easily have 
	\begin{align}
	&\mathcal{P}_{c, k}(\theta_c) = \mathbb{P}\left(\text{SINR}_U \geq \theta_c | \mathbf{b}_0 \in \Phi_k\right) \nonumber\\
	&=\pi \Xi \int_{0}^{\infty} \exp\Big(-\frac{2 \pi v \theta_c \Xi}{\alpha - 2}{}_2F_1(1, 1 - \frac{2}{\alpha}; 2 - \frac{2}{\alpha}; \theta_c)\Big) \nonumber\\
	&\quad \times \exp\Big(- \frac{\pi  v(\theta_c P_U)^{2 / \alpha} \sum_{t \in \mathcal{K}} \lambda_t}{\sinc(2 / \alpha)}\Big) \nonumber \\
	&\quad \times \exp\Big(-\frac{\pi v (\theta_c P_J)^{2 / \alpha} \lambda_J}{\sinc(2 / \alpha)}\Big) \nonumber\\
	&\quad \times \exp(-\pi v \Xi ) dv,
	\end{align}
	which is independent on the index $k$.
	From $\mathcal{P}_{c}(\theta_c) = \sum_{k \in \mathcal{K}} \mathcal{A}_k \mathcal{P}_{c, k}(\theta_c)$ and $\sum_{k \in \mathcal{K}} \mathcal{A}_k = 1$, we obtain $\mathcal{P}_c(\theta_c) = \mathcal{P}_{c, k}(\theta_c) \sum_{k \in \mathcal{K}} \mathcal{A}_k = \mathcal{P}_{c, k}(\theta_c)$.
	This gives the result in \eqref{eqn_user_connection_probability_simplified}.
\end{IEEEproof}
The user connection probability is now independent of the tier index.
It means that all users have the same connection performance in each tier.
In Sec. \ref{sec_numerical_results}, one can see that, although Lemma \ref{corollary_user_connection_probability_simplified} is developed based on several approximations, this property is almost true even without these approximations.

\subsection{User Secrecy Probability}
We investigate the secrecy probability of a randomly located user.
In this work, the user secrecy probability corresponds to the probability that a secret message for the typical user can not be decoded by any non-colluding eavesdroppers.
Hence the secrecy probability is defined as
\begin{equation}
\mathcal{P}_s(\theta_s, \tau) \triangleq \mathbb{P}(\bigcap_{\mathbf{e}_i \in \Phi_E}\text{SINR}_E(\mathbf{e}_i) < \theta_s ),
\end{equation}
where $\text{SINR}_E$ is given by 
\begin{equation}
\label{eqn_eavesdropper_SINR}
\text{SINR}_E(\mathbf{e}_i) = \frac{P_{\mathbf{b}_0} h_{\mathbf{b}_0, \mathbf{e}_i} D_{\mathbf{b}_0, \mathbf{e}_i}^{-\alpha}}{I'_B + I'_U + I'_J + N_0}.
\end{equation}
A randomly located eavesdropper is denoted as $\mathbf{e}_i$.
The cumulative interference from all the BSs (except the tagged BS) is given by $I'_B = \sum_{t \in \mathcal{K}}\sum_{\mathbf{b}_i \in \Phi_t \setminus \mathbf{b}_0} P_t h_{\mathbf{b}_i, \mathbf{e}_i} D_{\mathbf{b}_i, \mathbf{e}_i}^{-\alpha}$.
Similarly, $I'_U$ and $I'_J$ are the cumulative interference from scheduled users and active jammers, respectively, which are given by $I'_U = \sum_{t \in \mathcal{K}}\sum_{\mathbf{u}_i \in \Phi_{U,t}^s}  P_U h_{\mathbf{u}_i, \mathbf{e}_i} D_{\mathbf{u}_i, \mathbf{e}_i}^{-\alpha}$ and $I'_J = \sum_{\mathbf{z}_i \in \Phi_J^s} P_J h_{\mathbf{z}_i, \mathbf{e}_i} D_{\mathbf{z}_i, \mathbf{e}_i}^{-\alpha}$.
In this subsection, the point process of the scheduled jammers is modeled as a \emph{standard} PHP, because the locations of the holes (i.e. the scheduled users) and those of the eavesdroppers are mutually independent.
Thus we adopt the approach proposed in \cite{yazdanshenasan_poisson_2016} and only the closest hole to the eavesdropper $\mathbf{e}_i$ is considered.

\begin{theorem}
\label{theorem_secrecy_probability}
The user secrecy probability of a typical user is given by
\begin{align}
\label{eqn_user_secrey_probability}
\mathcal{P}_s(\theta_s, \tau) &= \sum_{k \in \mathcal{K}}\mathcal{A}_k \exp \bigg( -2 \pi \lambda_E \int_{0}^{\infty} \exp(-N_0 \gamma_{k}(r)) \nonumber\\
&\times \mathcal{L}_{I'_B}(\gamma_{k}(r)) \mathcal{L}_{I'_U}(\gamma_{k}(r)) \mathcal{L}_{I'_J}(\gamma_{k}(r)) r dr\bigg),
\end{align}
where $\gamma_{k}(r) = \frac{\theta_s r^{\alpha}}{P_k}$,
\begin{align}
\label{eqn_Laplace_BS2eav}
\mathcal{L}_{I'_B}(s) =& \exp\Big(- \frac{\pi s^{2 / \alpha} \sum_{t \in \mathcal{K}} \lambda_t P_t^{2 / \alpha}}{\sinc(\frac{2}{\alpha})}  \Big),\\
\label{eqn_Laplace_user2eav}
\mathcal{L}_{I'_U}(s) =& \exp \Big(-  \frac{ \pi \lambda_U^{s'} (s P_U)^{2 / \alpha} }{\sinc(\frac{2}{\alpha})} \Big), \\
\label{eqn_Laplace_jammer2eav}
\mathcal{L}_{I'_J}(s) =& \exp\Big( - \frac{\pi \lambda_J (s P_J)^{2 / \alpha}}{\sinc(\frac{2}{\alpha})} \Big) \nonumber\\ 
&\times \int_{0}^{\infty} G(v, s) 2 \pi \lambda_U^{s'} \exp(-\pi \lambda_U^{s'} v^2) v dv,
\end{align}
where $\lambda_U^{s'} = \sum_{t \in \mathcal{K}} \lambda_t$, and 

\begin{align}
	&G(v, s) = \nonumber\\
	&\begin{cases}
		\exp\big( \int_{v - R_{\tau}}^{v + R_{\tau}} \frac{2 y\lambda_J'(y)}{1 + \frac{y^{\alpha}}{s P_J}} dy \big), v > R_{\tau},\\
		\exp(\pi \lambda_J (R_{\tau} - v)^2 {}_2F_1(1, \frac{2}{\alpha}; 1 + \frac{2}{\alpha}; - \frac{(R_{\tau} - v)^{\alpha}}{s P_J}))\\ \quad\times \exp\big( \int_{R_{\tau} - v}^{R_{\tau} + v} \frac{2 y \lambda_J'(y)}{1 + \frac{y^{\alpha}}{s P_J}} dy \big), v \leq R_{\tau}.
	\end{cases}
\end{align}
\end{theorem}
\begin{IEEEproof}
Please refer to Appendix \ref{appendix_secrecy_probability}.
\end{IEEEproof}

It is worth mentioning that, although we adopt a few approximations in this section, our theoretical analyses show a good agreement with the simulation results in Sec. \ref{sec_numerical_results}.

\section{User Connection Probability Maximization}
\label{sec_maximization}
In the previous section, we derive the user connection probability and the user secrecy probability for given SINR thresholds $\theta_c$, $\theta_s$ and jammer selection threshold $\tau$.
However, in a practical system, it is necessary to design these thresholds according to the users' requirements.
In this section, we try to optimize these thresholds according to the users' quality of services (QoSs) in security.

\subsection{Problem Formulation}
To setup a secrecy transmission, there are two critical requirements: the secrecy probability $\mathcal{P}_s(\theta_s, \tau)$ and the secrecy rate $\mathcal{R}_s$.
The secrecy probability is required for transmitting messages securely since locations and CSI of the passive eavesdroppers are unknown.
A given secrecy rate is required for designing the secure coding and channel coding.
The secrecy rate is defined as $\mathcal{R}_s \triangleq [\mathcal{R}_U(\theta_c) - \mathcal{R}_E(\theta_s)]^+$, where $\mathcal{R}_U(\theta_c) = \log_2(1 + \theta_c)$ and $\mathcal{R}_E(\theta_s) = \log_2(1 + \theta_s)$ are the codeword rate and the redundant rate, respectively.
Subject to these QoSs, the user connection probability should be as large as possible.
Hence, we study the optimization problem as 
\begin{subequations}
	\label{eqn_optimization_problem}
	\begin{align}
	\max_{\theta_c, \theta_s, \tau} \quad&\mathcal{P}_c(\theta_c, \tau), \\
	\text{s.t.} \quad&\mathcal{P}_s(\theta_s, \tau) \geq \mathcal{P}_T,  \\
	& \mathcal{R}_U(\theta_c) - \mathcal{R}_E(\theta_s) \geq \mathcal{R}_T,
	\end{align}
\end{subequations}
where $\mathcal{P}_T$ and $\mathcal{R}_T$ are the requirements of secrecy probability and secrecy rate, respectively.

Plugging the secrecy rate definition into \eqref{eqn_optimization_problem}, the problem can be rewritten as
\begin{subequations}
	\begin{align}
	\max_{\theta_c, \theta_s, \tau} \quad&\mathcal{P}_c(\theta_c, \tau), \\
	\text{s.t.} \quad&\mathcal{P}_s(\theta_s, \tau) \geq \mathcal{P}_T,  \\
	& \theta_c \geq 2^{\mathcal{R}_T}(1 + \theta_s) - 1.
	\end{align}
\end{subequations}
According to the user connection probability definition \eqref{eqn_user_connection_probability_definition}, the objective function is monotonically decreasing with $\theta_c$.
Hence the optimal solution would be obtained when $\theta_c = 2^{\mathcal{R}_T}(1 + \theta_s) - 1$, and the problem could be rewritten as
\begin{subequations}
	\label{eqn_optimization_problem_2}
	\begin{align}
	\max_{\theta_s, \tau} \quad&\mathcal{P}_c(\theta_c, \tau), \\
	\text{s.t.} \quad&\mathcal{P}_s(\theta_s, \tau) \geq \mathcal{P}_T, \\
	& \theta_c = 2^{\mathcal{R}_T}(1 + \theta_s) - 1.
	\end{align}
\end{subequations}
Moreover, one can see that the user connection probability and the user secrecy probability have inverse monotone.
The user connection probability is monotonically decreasing with $\theta_s$ (and $\theta_c$), while the secrecy probability is monotonic increasing.
As the threshold $\tau$ increases, more jammers are selected to be active and the secrecy probability increases.
On the other hand, as more jammers transmitting artificial noise, the user's SINR performance degrades and $\mathcal{P}_c(\theta_c, \tau)$ decreases.
Thus the optimal solution would be obtained when $\mathcal{P}_s(\theta_s, \tau) = \mathcal{P}_T$, and the problem \eqref{eqn_optimization_problem_2} can be rewritten as
\begin{subequations}
	\label{eqn_optimization_problem_3}
	\begin{align}
	\max_{\theta_s, \tau} \quad&\mathcal{P}_c(\theta_c, \tau), \\
	\label{eqn_optimization_problem_3_constraint_1}
	\text{s.t.} \quad&\mathcal{P}_s(\theta_s, \tau) = \mathcal{P}_T, \\
	& \theta_c = 2^{\mathcal{R}_T}(1 + \theta_s) - 1.
	\end{align}
\end{subequations}

\subsection{Greedy Algorithm}
As shown in Theorem \ref{theorem_connection_probability} and Theorem \ref{theorem_secrecy_probability}, the expressions of the user connection/secrecy probability are very complicated.
Hence it is difficult to obtain the optimal solution of the problem \eqref{eqn_optimization_problem_3}.
In this subsection, we propose a greedy algorithm to solve the optimization problem efficiently, which is presented in Algorithm \ref{alg_thresholds_optimization}.

\begin{algorithm}
	\centering
	\caption{Thresholds Optimization Algorithm}
	\label{alg_thresholds_optimization}
	\begin{algorithmic}[1]
		\STATE Input: $\mathcal{P}_T$, $\mathcal{R}_T$, and $\Delta$;
		\STATE Initialization: $R_{\tau}^{(0)} = 0$, $\theta_c^{(0)} = 0$, $\theta_s^{(0)} = 0$;
		\LOOP
			\STATE Calculate $\tau^{(t)} = \frac{P_J}{(R_{\tau}^{(t)})^{\alpha}};$
			\STATE Using the bisection method, calculate $\theta_s^{(t)}$ that satisfies $\mathcal{P}_s(\theta_s^{(t)}, \tau^{(t)}) = \mathcal{P}_T$;
			\STATE Calculate $\theta_c^{(t)} = 2^{\mathcal{R}_T} (1 + \theta_s^{(t)}) - 1$;
			\IF {$\mathcal{P}_c(\theta_c^{(t)}, \tau^{(t)}) > \mathcal{P}_c(\theta_c^{(t - 1)}, \tau^{(t - 1)})$}
				\STATE $R_{\tau}^{(t + 1)} =  R_{\tau}^{(t)} + \Delta$;
			\ELSE
				\RETURN $\theta_s^* = \theta_s^{(t - 1)}$, $\theta_c^* = \theta_c^{(t - 1)}$ and $\tau^* = \tau^{(t - 1)}$.
			\ENDIF
		\ENDLOOP
	\end{algorithmic}
\end{algorithm}
Algorithm \ref{alg_thresholds_optimization} begins with no jammer selection, i.e. $R_{\tau}^{(0)} = 0$ ($\tau^{(0)} = \infty$) (Line 2).
In each iteration, we try to increase the user connection probability by increasing $R_{\tau}$ (Line 7 to 8), since $\mathcal{P}_c$ is a monotonically increasing function of $R_{\tau}$ (Line 4 is the relationship between $\tau$ and $R_{\tau}$).
$\Delta$ is the step size of $R_{\tau}$.
Due to the complicated expression of $\mathcal{P}_s$ in \eqref{eqn_user_secrey_probability}, it is difficult to derive an expression for $\theta_s^{(t)}$ from $\mathcal{P}_s(\theta_s^{(t)}, \tau^{(t)}) = \mathcal{P}_T$.
However, we can efficiently calculate $\theta_s^{(t)}$ that satisfies $\mathcal{P}_s(\theta_s^{(t)}, \tau^{(t)}) = \mathcal{P}_T$ using the bisection method (Line 5), since $\mathcal{P}_s(\theta_s, \tau)$ is a monotonically increasing function of $\theta_s$.
The algorithm ends when there is no improvement in the user connection probability and returns $\theta_s^*$, $\theta_c^*$ and $\tau^*$ (Line 9 to 10).

\begin{lemma}
	\label{lemma_optimization}
	Given $\tau$, the output of Algorithm \ref{alg_thresholds_optimization} is optimal.
\end{lemma}
\begin{IEEEproof}
	The user secrecy probability $\mathcal{P}_s(\theta_s, \tau)$ is monotonic with $\theta_s$.
	Hence, given $\tau$, there is only one root $\theta_s^*$ satisfying the constraint \eqref{eqn_optimization_problem_3_constraint_1}, which can be obtained by using the bisection method in Algorithm \ref{alg_thresholds_optimization}.
\end{IEEEproof}

\begin{figure}[t]
	\centering
	\includegraphics[width=85mm]{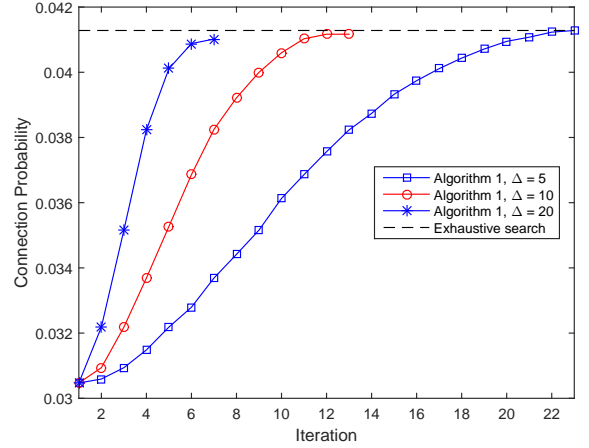}
	\vspace{-0.1cm}
	\caption{The iteration process and the performance of Algorithm 1 with different step sizes}
	\label{fig_optimization_process}
	\vspace{-0.3cm}
\end{figure}

As shown in Lemma \ref{lemma_optimization}, the output of Algorithm \ref{alg_thresholds_optimization} in each iteration is optimal.
Moreover, the objective function increases in each iteration.
Hence the proposed algorithm converges quickly.
In Fig. \ref{fig_optimization_process}, we show three examples of the iteration process of Algorithm \ref{alg_thresholds_optimization} with different step sizes $\Delta$.
The system parameters are the same as those in Sec. \ref{sec_numerical_results}.
One can see that, the proposed algorithm converges quickly and approaches to the performance of exhaustive search.
The step size $\Delta$ has a great impact on the number of iteration.
A larger step size can speed up the convergence but might lead to a suboptimal result (as shown by the curve with star marks).

\section{Numerical Results}
\label{sec_numerical_results}
In this section, we evaluate the theoretical analyses and demonstrate the effectiveness of our proposed optimization algorithm by using numerical results.
A two-tier HetNet is simulated.
The densities of two tiers are $\lambda_1 = 1 \text{ BS}/\text{km}^2$ and $\lambda_2 = 10 \text{ BS} / \text{km}^2$. 
The user density is $\lambda_U = 100 \text{ user} / \text{km}^2$. 
We assume that there is one eavesdropper around each BS on average, i.e. $\lambda_E = 11 \text{ eavesdropper} / \text{km}^2$.
The jammer density is $\lambda_J = 50 \text{ jammer} / \text{km}^2$. 
The path loss exponent $\alpha$ is assumed to be $3.5$. 
The transmit power are set as follows: $P_1 = 46$ dBm, $P_2 = 30$ dBm, $P_U = 23$ dBm and $P_J = 23$ dBm, respectively.
The jammer selection threshold $\tau$ is $-80$ dB.
We assume that the SIC capacity is $\beta = -90$ dB and the thermal noise power is $N_0 = - 174$ dBm.
All the results in this section are obtained under the above parameter settings unless noted otherwise.

\subsection{Validation and Insights of the Theoretical Analyses}
\begin{figure*}[t]
	\centering
	\includegraphics[width=160mm]{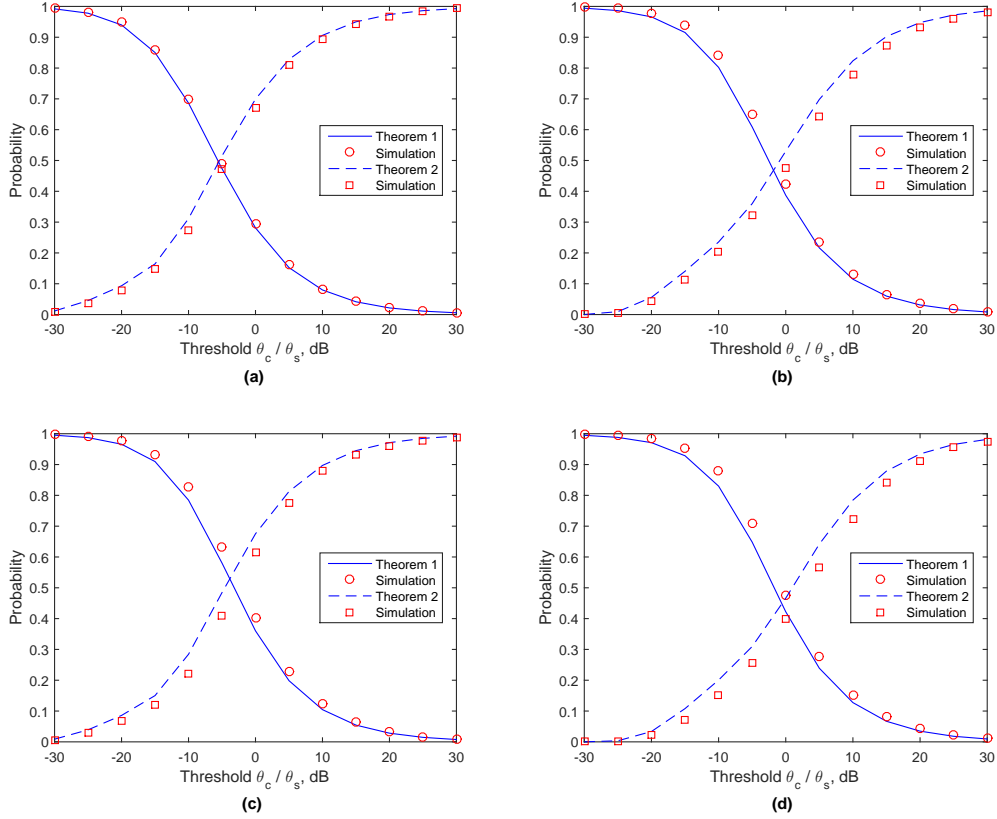}
	\vspace{-0.1cm}
	\caption{Analytical and simulation results for the user connection/secrecy probability in four configurations: (a) LD-SH, (b) HD-SH, (c) LD-LH, (d) HD-LH.}
	\label{fig_SINR_simulation_VS_theorem}
	\vspace{-0.3cm}
\end{figure*}

Since we only consider the two closest holes of active jammers in Theorem \ref{theorem_connection_probability} (the closest hole in Theorem \ref{theorem_secrecy_probability}), our analyses can provide tighter bounds on the simulation results where the holes are small and sparse compared with the system configuration where the holes are large and dense.
Therefore, we define four possible configurations: 
low density of holes and small holes (LD-SH), high density of holes and small holes (HD-SH), low density of holes and large holes (LD-LH), and high density of holes and large holes (HD-LH).
As only one user per BS interferes with the typical user, the density of the holes is determined by the densities of BSs.
The system parameters of the four configurations are summarized in Table \ref{table_configuration}.
Fig. \ref{fig_SINR_simulation_VS_theorem} shows the user connection/secrecy probabilities in the four configurations given by Theorem \ref{theorem_connection_probability} and Theorem \ref{theorem_secrecy_probability} along with the Monte Carlo simulation results.
One can observe that, our analytical results agree well with the simulations in four cases.
Theorem \ref{theorem_connection_probability} and Theorem \ref{theorem_secrecy_probability} are the lower and upper bounds respectively, since the derived Laplace transform of interference from active jammers is a lower bound on the accurate results.

\begin{table}[t]
	\renewcommand{\arraystretch}{1.3}
	\caption{System Parameters Summary of Four Configurations}
	\label{table_configuration}
	\centering
	\begin{tabular}{|c|c|c|c|}
		\hline
		\bfseries Configuration & BS densities ($\text{BS}/\text{km}^2$) & $\tau$ (dB) & Hole radius (m)\\
		\hline
		\hline
		LD-SH & $\lambda_1 = 1, \lambda_2 = 2$ & $-77$ & $R_{\tau} \approx 100$ \\
		\hline
		HD-SH & $\lambda_1 = 2, \lambda_2 = 10$ & $-77$ & $R_{\tau} \approx 100$\\
		\hline
		LD-LH & $\lambda_1 = 1, \lambda_2 = 2$ & $-88$ & $R_{\tau} \approx 200$\\
		\hline
		HD-LH & $\lambda_1 = 2, \lambda_2 = 10$ & $-88$ & $R_{\tau} \approx 200$\\
		\hline
	\end{tabular}
\end{table}

\begin{figure}[t]
	\centering
	\includegraphics[width=85mm]{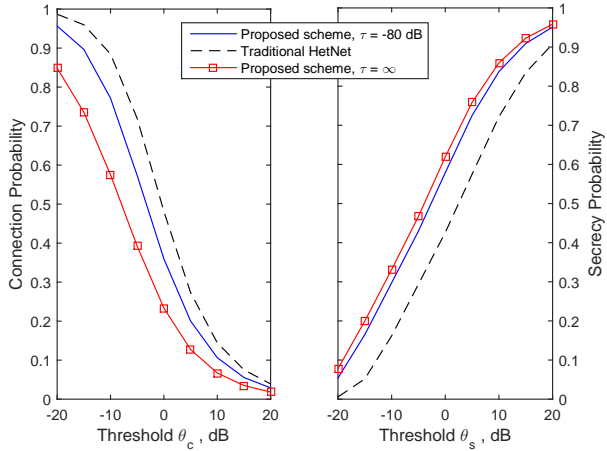}
	\vspace{-0.1cm}
	\caption{The user connection/secrecy performance comparison among (i) the proposed scheme, (ii) the traditional HetNet, and (iii) the HetNet in which all the jammers and the scheduled users transmit artificial noise. }
	\label{fig_performance_comparison}
	\vspace{-0.3cm}
\end{figure}
In Fig. \ref{fig_performance_comparison}, our proposed scheme is compared with two benchmarks: (i) the traditional HetNet in which there is neither jammer nor user transmitting artificial noise; and (ii) the HetNet in which all the jammers and the scheduled users transmit artificial noise, i.e. $\tau = \infty$.
As expected, our scheme provides a good balance between the user connection probability and the user secrecy probability.
The benchmark (ii) provides the best security performance, because the interference level in it is the strongest.
One can see that our scheme provides a close secrecy probability to it.
Meanwhile, the proposed scheme and the traditional HetNet have similar performance in the user connection probability, which is much better than that of the benchmark (ii).

\begin{figure}[t]
	\centering
	\includegraphics[width=85mm]{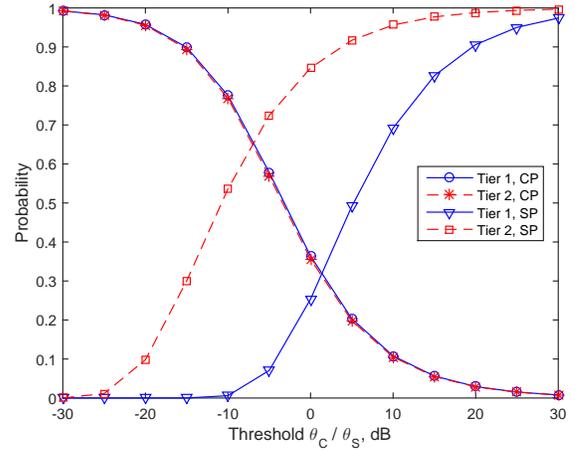}
	\vspace{-0.1cm}
	\caption{The user connection/secrecy probability in the two tiers.}
	\label{fig_performance_in_each_tier}
	\vspace{-0.3cm}
\end{figure}
In Fig. \ref{fig_performance_in_each_tier}, we show the user connection probability (CP) and secrecy probability (SP) in the two tiers respectively.
Although here we do not adopt the approximations used in deriving Corollary \ref{corollary_user_connection_probability_simplified}, the user connection probabilities in the two tiers are still quite similar. 
Hence, maximizing the probability $\mathcal{P}_c(\theta_c, \tau)$ can provide the best connection performance to the users in each tier simultaneously.
On the other hand, the user secrecy probability in tier $2$ is much better than that in tier $1$.
It is because the transmit power of BSs in tier $2$ is much lower.
The associated users in tier $2$ are located closely to their serving BSs, while the eavesdroppers are suffering from low received power.
Hence the tier with lower transmit power would be better choice for secrecy wireless communication.
Furthermore, the secrecy performance gap inspires that the system performance with unequal QoS requirements is worth studying in the future.

\subsection{System Performance with the Threshold Optimization Algorithm}
In this subsection, we show the output of problem \eqref{eqn_optimization_problem} obtained by Algorithm \ref{alg_thresholds_optimization}.
The requirements of the secrecy probability and the secrecy rate are assumed to be $\mathcal{P}_T = 0.9$ and $\mathcal{R}_T = 1$ bps, respectively.
The thresholds $\theta_c$, $\theta_s$ and $\tau$ are computed by Algorithm \ref{alg_thresholds_optimization} with $\Delta = 5$.

\begin{figure}[t]
	\centering
	\includegraphics[width=85mm]{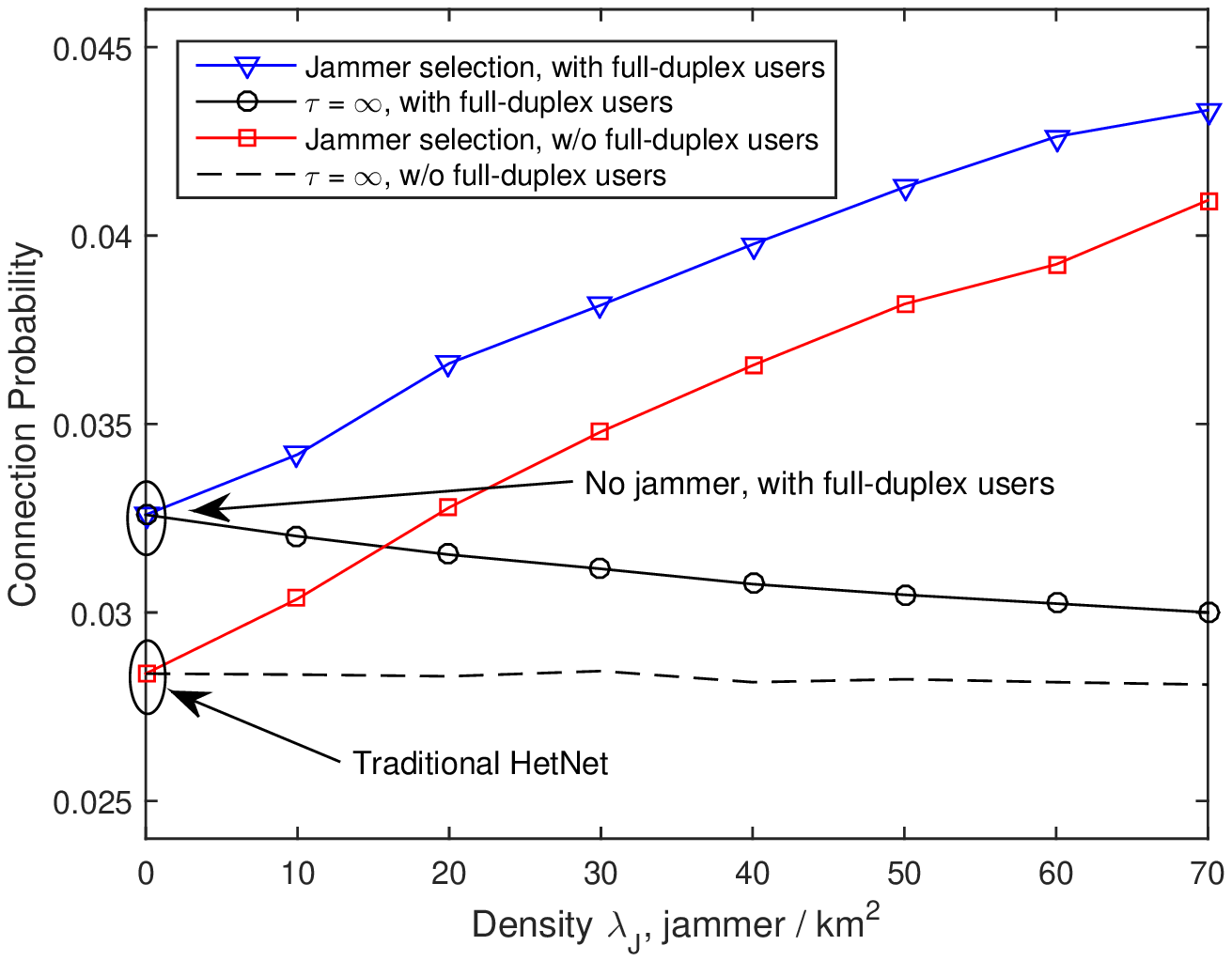}
	\vspace{-0.1cm}
	\caption{Connection performance obtained by Algorithm \ref{alg_thresholds_optimization} with varying jammer density.}
	\label{fig_optimization_result_VS_jammer_density}
	\vspace{-0.3cm}
\end{figure}
\begin{figure}[t]
	\centering
	\includegraphics[width=85mm]{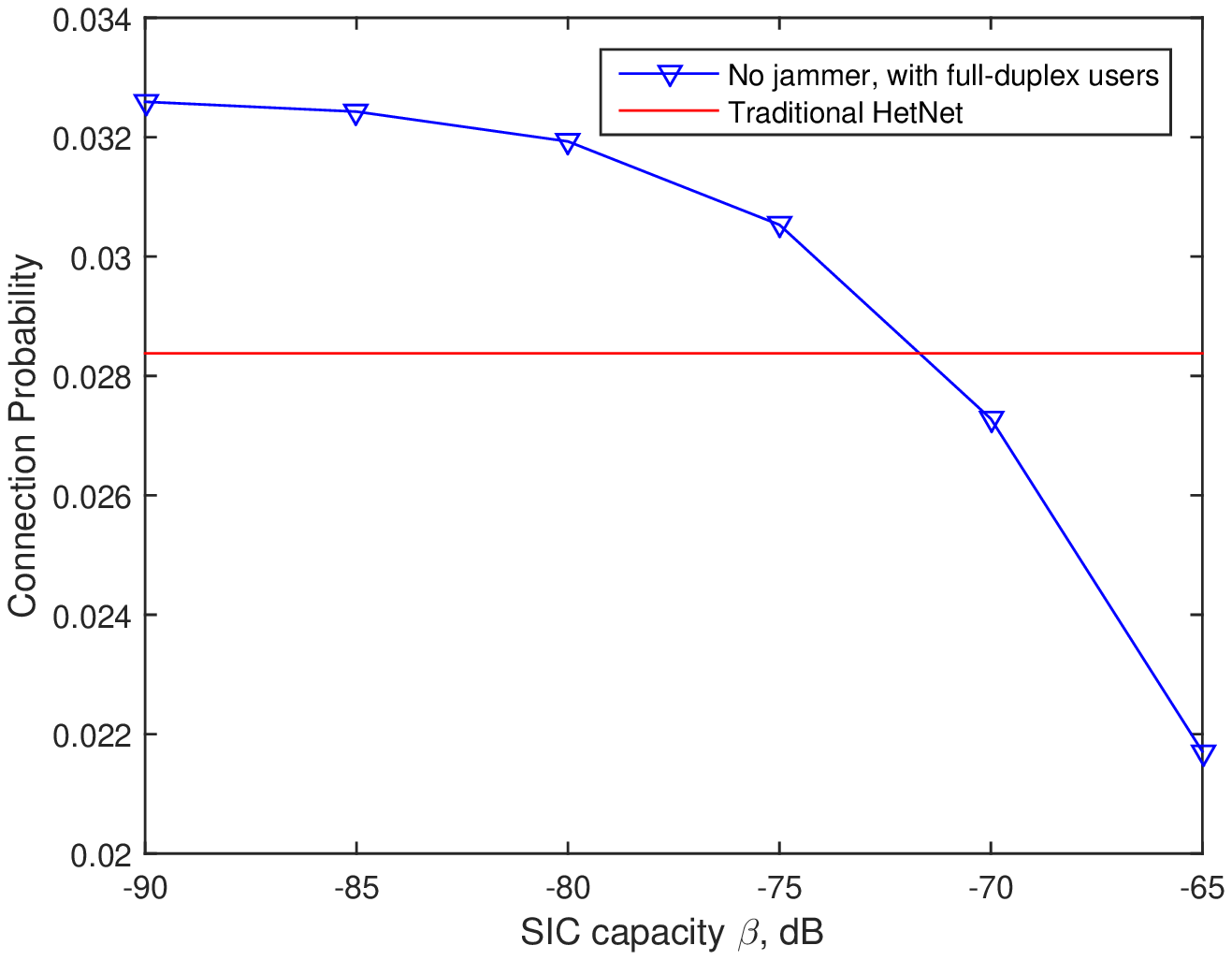}
	\vspace{-0.1cm}
	\caption{Connection performance obtained by Algorithm \ref{alg_thresholds_optimization} with varying SIC capacity.}
	\label{fig_optimization_result_VS_SIC_capability}
	\vspace{-0.3cm}
\end{figure}
The performance of different schemes with varying jammer density are shown in Fig. \ref{fig_optimization_result_VS_jammer_density}.
One can see that simply deploying more jammers can not improve the system performance.
Without jammer selection, the connection probability gets worse when the density of jammers grows (i.e. the solid curve with circle marks and the dashed curve).
On the other hand, the schemes with jammer selection (i.e. the curves with triangle and square marks) significantly outperform those without jammer selection.
The performance gaps increase with the jammer density.
As shown in Fig. \ref{fig_performance_comparison}, the proposed scheme can provide a good connection probability while maintaining a high secrecy performance.
Hence, under the same security requirements, our scheme provides much better connection performance.
According to the jammer selection policy, the jammers whose interference on any scheduled user is stronger than $\tau$ are prevented from being active.
When the jammer density increases, there are more jammer located closely to the eavesdroppers, but the interference level suffered by the users from jammers is mitigated and maintains low.
Hence the performance in the schemes with jammer selection increases with the density of jammer. 
Furthermore, transmitting artificial noise by scheduled full-duplex users also improves the system performance remarkably, which is shown by the gap between the curves with triangle and square marks.
When there is no jammer (i.e. $\lambda_J = 0$), the scheme with full-duplex users provides a $15 \%$ gain in the connection probability compared with the traditional HetNet.
However, this performance gain evidently depends on the SIC capacity $\beta$.
Fig. \ref{fig_optimization_result_VS_SIC_capability} shows the performance in the two schemes with varying SIC capacity.
One can see that the SIC capacity has significant impact on the connection probability.
When $\beta$ is larger than $-70$ dB, there is no performance gain obtained by letting full-duplex scheduled users transmit artificial noise. 

\begin{figure}[t]
	\centering
	\includegraphics[width=85mm]{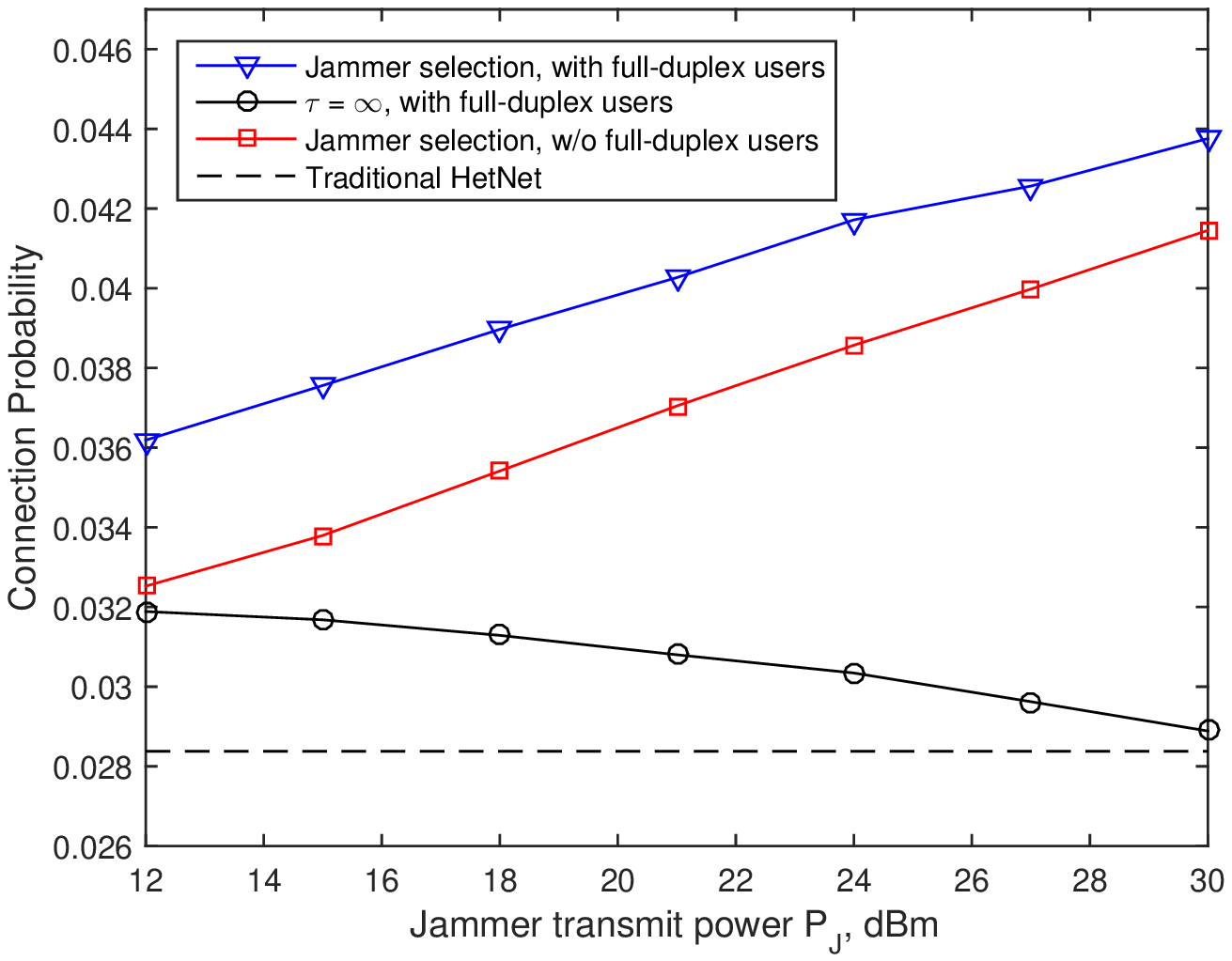}
	\vspace{-0.1cm}
	\caption{Connection performance obtained by Algorithm \ref{alg_thresholds_optimization} with varying jammer transmit power.}
	\label{fig_optimization_result_VS_jammer_transmit_power}
	\vspace{-0.3cm}
\end{figure}
We show the performance of different schemes with varying jammer transmit power in Fig. \ref{fig_optimization_result_VS_jammer_transmit_power}.
Simply increasing the transmit power of jammer without jammer selection degrades the system performance (as shown by the curve with circle marks).
On the other hand, the schemes with jammer selection provide remarkable connection probability gain.
The reason is similar to that of Fig. \ref{fig_optimization_result_VS_jammer_density}.
Without the jammer selection policy, the interference level suffered by the users increases with the jammer transmit power.
On the other hand, the interference to the scheduled users could be mitigated with jammer selection, while the received SINRs of eavesdroppers drop greatly when $P_J$ increases.
Hence our scheme could provide significant improvement under the same QoS requirements. 

\begin{figure}[t]
	\centering
	\includegraphics[width=85mm]{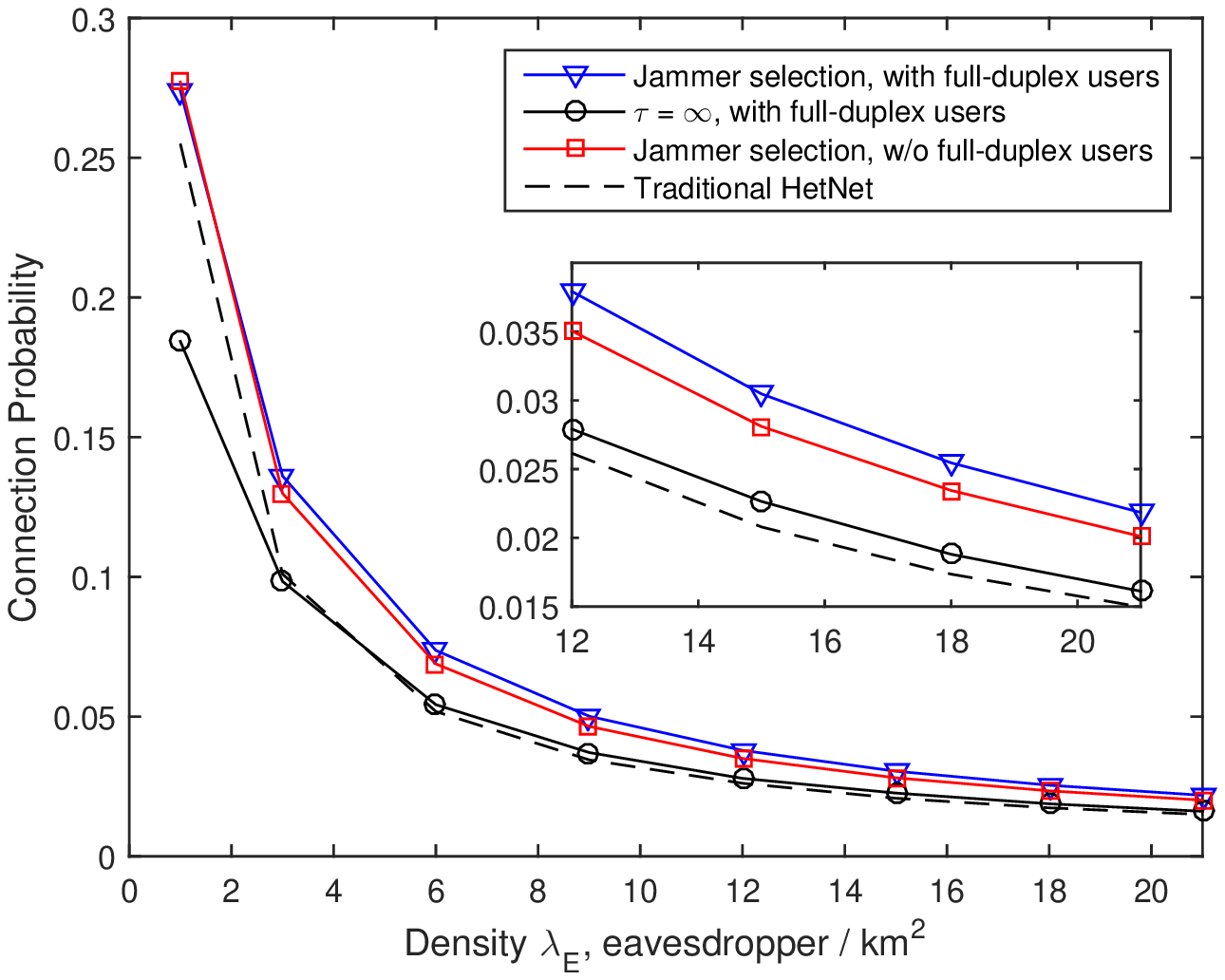}
	\vspace{-0.1cm}
	\caption{Connection performance obtained by Algorithm \ref{alg_thresholds_optimization} with varying eavesdropper density.}
	\label{fig_optimization_result_VS_eavesdropper_density}
	\vspace{-0.3cm}
\end{figure}
Fig. \ref{fig_optimization_result_VS_eavesdropper_density} shows the performance of different schemes with varying eavesdropper density.
As expected, the eavesdropper density has a great impact on the system performance.
The schemes with jammer selection provide the best performance with the density varying from $1$ to $21$  $\text{eavesdropper} / \text{km}^2$.
However, it is interesting to see that, when the density is small (e.g. $\lambda_E = 1 \text{ eavesdropper} / \text{km}^2$), the schemes without full-duplex users provide better performance than those with full-duplex users (the curve with square marks vs the one with triangle marks, the dashed curve vs the solid curve with circle marks). 
When the eavesdropper density is sufficiently small, there are few eavesdroppers located near the scheduled users.
Hence, it might not be worth confounding the eavesdroppers using full-duplex given the cost of suffering self-interference and interference from other scheduled users.

\section{Conclusion}
\label{sec_conclusion}
This paper proposed a novel HetNet scheme with jammer selection and full-duplex users.
We characterized the user connection probability and user secrecy probability by using stochastic geometry.
Especially, inhomogeneous PPP and (modified) PHP were used for modeling the scheduled users and the active jammers, respectively.
A greedy algorithm for jointly optimizing the SINR thresholds for user connection, secrecy transmission and the jammer selection threshold subject to security QoSs was also proposed.
Extensive performance evaluations were conducted.
The results showed that the proposed HetNet scheme could improve the system performance significantly. 

\section*{Acknowledgment}
The authors would like to thank the anonymous reviewers.
Their valuable suggestions greatly improve the quality of this paper.

\appendices
\section{}
\label{appendix_scheduled_user_intensity}
\begin{figure}[t]
	\centering
	\includegraphics[width=85mm]{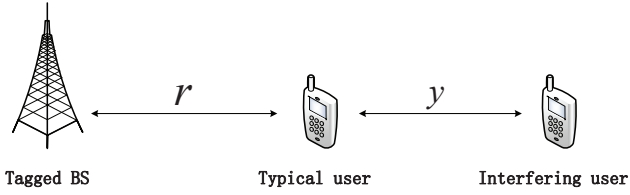}
	\vspace{-0.1cm}
	\caption{An instance of the distance between nodes.}
	\label{fig_Scheduled_User_Intensity}
	\vspace{-0.3cm}
\end{figure}

We have \eqref{eqn_scheduled_user_intensity_tier_t} straightly following the derivation of (21) in \cite{tang_hybrid_2016}. 
For the integrity of content, we show the derivation here.
As shown in Fig. \ref{fig_Scheduled_User_Intensity}, $r$ denotes as the distance between the typical user and its serving BS, and $y$ is the distance from the interfering user to the typical user.
A user at distance $r + y$ is associated with a BS in tier $t$ but not the tagged BS with probability $1 - \exp\big(-\pi (r+y)^2 \lambda_t (\frac{P_t}{P_k})^{2 / \alpha}\big)$. 
Note that, in each slot, there is always only one user scheduled in each BS.
We have the intensity of the interfering users in tier $t$ as 
\begin{equation}
\label{eqn_intensity_interfering_user}
\lambda_{U,t}^s(r, y) = \lambda_t \Big(1 - \exp(-\pi (r+y)^2 \lambda_t (\frac{P_t}{P_k})^{2 / \alpha})\Big).
\end{equation} 
Using the property of Poisson process \cite{baccelli_stochastic_2009}, we have the intensity measure function \eqref{eqn_scheduled_user_intensity}.

It is worth mentioning that, \eqref{eqn_intensity_interfering_user} is not a mathematical approximation of the intensity measure of Poisson-Voronoi perturbed lattice but a tractable modeling method.
Compared with the homogeneous PPP \cite{elsawy_stochastic_2014, lee_hybrid_2015, liu_tractable_2016}, the inhomogeneous PPP captures the correlation among the scheduled users and is more accurate \cite{singh_joint_2015}.
This is validated in \cite{singh_joint_2015, tang_hybrid_2016} and Sec. \ref{sec_numerical_results}.

\section{}
\label{appendix_laplace_jammer2user}
\begin{figure}[t]
	\centering
	\includegraphics[width=85mm]{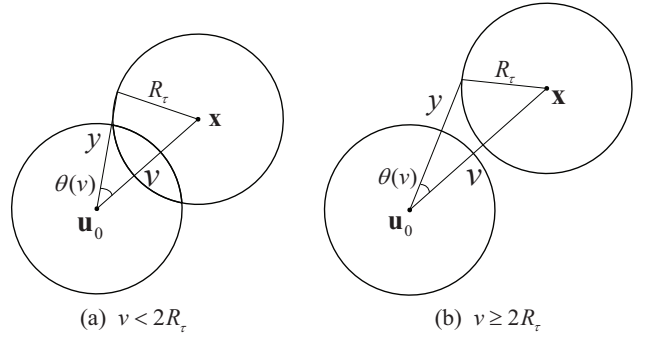}
	\vspace{-0.1cm}
	\caption{Illustration of $\mathbf{B}^c(\mathbf{u}_0, R_{\tau}) \bigcap \mathbf{B}(\mathbf{x}, R_{\tau})$.}
	\label{fig_Closest_Two_Holes}
	\vspace{-0.3cm}
\end{figure}

The Laplace transform of the interference from active jammers is
\begin{align}
\mathcal{L}_{I_J}(s) &= \mathbb{E} \Big[\exp(-s(\sum_{\mathbf{z}_i \in \Phi_J^s} P_J h_{\mathbf{z}_i, \mathbf{u}_0} D_{\mathbf{z}_i, \mathbf{u}_0}^{-\alpha})) \Big] \nonumber\\
&\overset{(a)}{=} \mathbb{E}_{\Phi_J^s} \Big[\prod_{\mathbf{z}_i \in \Phi_J^s} \mathcal{L}_h(s P_J D_{\mathbf{z}_i, \mathbf{u}_0}^{-\alpha}) \Big],
\end{align}
where $(a)$ follows that the small-scale fading on different nodes is i.i.d. $h \sim \exp(1)$ and mutually independent with $\Phi_J^s$.
According to our approach, we consider only two holes: (i) the one around the typical user, and (ii) the one that is closest to the typical user, and ignore the other holes.
Hence, the Laplace transform can be expressed as
\begin{align}
\mathcal{L}_{I_J}(s) &\approx \mathbb{E}_{\mathbf{x}}\Big[ \mathbb{E}_{\Phi_J} \Big[ \prod_{\mathbf{z}_i \in \Phi_J \setminus \mathbf{C}} \mathcal{L}_h(s P_J D_{\mathbf{z}_i, \mathbf{u}_0}^{-\alpha}) \Big]\Big] \nonumber\\
&\overset{(a)}{=} \mathbb{E}_{\mathbf{x}} \Big[ \exp\Big(-\lambda_J \int\limits_{\mathbb{R}^2 \setminus \mathbf{C}}(1 - \mathcal{L}_h(s P_J \|\mathbf{z}\|^{-\alpha})) d\mathbf{z}\Big) \Big],
\end{align}
where $\mathbf{C} = \mathbf{B}(\mathbf{u}_0, R_{\tau}) \bigcup \mathbf{B}(\mathbf{x}, R_{\tau}) = \mathbf{B}(\mathbf{u}_0, R_{\tau}) \bigcup \Big(\mathbf{B}^c(\mathbf{u}_0, R_{\tau}) \bigcap \mathbf{B}(\mathbf{x}, R_{\tau})\Big)$.
$\mathbf{B}^c(\mathbf{u}_0, R_{\tau})$ is the complementary set of $\mathbf{B}(\mathbf{u}_0, R_{\tau})$.
$\mathbf{x}$ is the location of the closest hole to the typical user and $(a)$ follows the PGFL of PPP.
Then we have
\begin{align}
\label{eqn_Laplace_Jammer_derivation}
&\mathcal{L}_{I_J}(s) \nonumber\\
&\approx \mathbb{E}_{\mathbf{x}} \Big[\exp\Big(-\lambda_J \int_{\mathbb{R}^2 \setminus \mathbf{B}(\mathbf{u}_0, R_{\tau})}(1 - \mathcal{L}_h(s P_J \|\mathbf{z}\|^{-\alpha})) d\mathbf{z} \nonumber\\
&\quad + \lambda_J \int_{\mathbf{B}^c(\mathbf{u}_0, R_{\tau}) \bigcap \mathbf{B}(\mathbf{x}, R_{\tau})}(1 - \mathcal{L}_h(s P_J \|\mathbf{z}\|^{-\alpha})) d\mathbf{z}\Big)\Big]\nonumber\\
&= \exp\Big(-2\pi \lambda_J \frac{R_{\tau}^{2 - \alpha} s P_J}{\alpha -2} {}_2F_1(1, 1 - \frac{2}{\alpha}; 2 - \frac{2}{\alpha}; \frac{-s P_J}{R_{\tau}^{\alpha}} )\Big) \nonumber\\
& \quad \times \mathbb{E}_{\mathbf{x}} \Big[\exp\Big(\lambda_J \int_{\mathbf{B}^c(\mathbf{u}_0, R_{\tau}) \bigcap \mathbf{B}(\mathbf{x}, R_{\tau})}\frac{1}{1 + \frac{\| \mathbf{z} \|^{\alpha}}{s P_J}} d\mathbf{z}\Big)\Big] \nonumber\\
&= \exp\Big(-2\pi \lambda_J \frac{R_{\tau}^{2 - \alpha} s P_J}{\alpha -2} {}_2F_1(1, 1 - \frac{2}{\alpha}; 2 - \frac{2}{\alpha}; \frac{-s P_J}{R_{\tau}^{\alpha}} )\Big) \nonumber\\
&\quad \times \int_{0}^{\infty} H(v, s)  f(v) dv,
\end{align}
where
\begin{align}
&H(v, s) = \exp\Big( \int_{\max\{v - R_{\tau}, R_{\tau}\}}^{v + R_{\tau}} \frac{2 y\lambda_J'(y)}{1 + \frac{y^{\alpha}}{s P_J}} dy \Big),\\
\label{eqn_f}
&f(v) = 2 \pi \lambda_U^s(r, v) v \exp(-2\pi \int_{0}^{v} \lambda_U^s(r, y) y dy),\\
&\lambda_J'(y) = \lambda_J\arccos(\frac{y^2 + v^2 - R_{\tau}^2}{2 y v}).
\end{align}
The first term in \eqref{eqn_Laplace_Jammer_derivation} is derived by converting from Cartesian to polar coordinates and the closed form expression follows from the properties of the Gamma function \cite[Appendix B]{jo_heterogeneous_2012}.
The second term follows from the cosine-law: $y^2 + v^2 - 2yv\cos \theta(y) = R_{\tau}^2$ (as shown in Fig. \ref{fig_Closest_Two_Holes}) and some geometry derivation. 
Note that the centers of holes are the scheduled users, and $\Phi_U^s$ is assumed to be inhomogeneous PPP with intensity measure function \eqref{eqn_scheduled_user_intensity}. 
Thus, the PDF of the distance $v \triangleq \|\mathbf{x}\|$ between the typical user at the origin and the closest hole $\mathbf{x}$ is given as \eqref{eqn_f}.

The integral $\int_{0}^{v} \lambda_{U}^s(r, y) y dy$ could be expressed in closed form as
\begin{align}
&\int_{0}^{v} \lambda_{U}^s(r, y) y dy \nonumber\\
&= \sum_{t \in \mathcal{K}}\Big(\frac{\lambda_t v^2}{2} - \lambda_t\Big(\frac{\exp(-\pi C_t r^2) - \exp(-\pi C_t (r + v)^2)}{2\pi C_t} \nonumber\\
&\qquad \qquad \quad- r\frac{\mathrm{erf}((r + v)\sqrt{\pi C_t}) - \mathrm{erf}(\sqrt{\pi C_t} r) }{2 \sqrt{C_t}}\Big) \Big),
\end{align}
where $C_t = \lambda_t (\frac{P_t}{P_k})^{2 / \alpha}$ and $\mathrm{erf}(x)$ is the error function.
For readability, we maintain the integral form in \eqref{eqn_Laplace_jammer2user}.

It is worth mentioning that, in our approach, we consider only two closest holes.
Hence \eqref{eqn_Laplace_jammer2user} is a lower bound on the accurate Laplace transform of the interference from active jammers.

\section{}
\label{appendix_secrecy_probability}
The user secrecy probability can be expressed as
\begin{align}
&\mathcal{P}_s(\theta_s, R_{\tau}) = \sum_{k \in \mathcal{K}} \mathcal{A}_k \mathbb{P}(\bigcap_{\mathbf{e}_i \in \Phi_E}\text{SINR}_E(\mathbf{e}_i) < \theta_s | \mathbf{b}_0 \in \Phi_k) \nonumber\\
&= \sum_{k \in \mathcal{K}} \mathcal{A}_k \mathbb{E}_{\Phi_E}\Big[\mathbb{P}(\bigcap_{\mathbf{e}_i \in \Phi_E}\text{SINR}_E(\mathbf{e}_i) < \theta_s | \mathbf{b}_0 \in \Phi_k, \Phi_E) \Big],
\end{align}
where $\text{SINR}_E(\mathbf{e}_i)$ is given by \eqref{eqn_eavesdropper_SINR}.
One can see that $\{D_{\mathbf{b}_0, \mathbf{e}_i}\}$ are correlated when conditioned on $\Phi_E$.
Hence, the probabilities $\Big\{\mathbb{P}(\text{SINR}_E(\mathbf{e}_i) < \theta_s | \mathbf{b}_0 \in \Phi_k, \Phi_E)\Big\}$ are not mutually independent.
For the sake of tractability, we ignore the correlations among these probabilities.
Hence, we have
\begin{align}
\label{eqn_appendix_c_1}
&\mathcal{P}_s(\theta_s, R_{\tau}) \nonumber\\
&\approx \sum_{k \in \mathcal{K}} \mathcal{A}_k \mathbb{E}_{\Phi_E} \Big[ \prod_{\mathbf{e}_i \in \Phi_E}  \mathbb{P}(\text{SINR}_E(\mathbf{e}_i) < \theta_s | \mathbf{b}_0 \in \Phi_k)\Big] \nonumber\\
&\overset{(a)}{=}\sum_{k \in \mathcal{K}} \mathcal{A}_k \mathbb{E}_{\Phi_E} \Big[\exp \Big(- \lambda_E \int_{\mathbb{R}^2} \mathbb{P}(\text{SINR}_E(\mathbf{e}) \geq \theta_s) d\mathbf{e}\Big) \Big],
\end{align}
where $(a)$ follows the PGFL of PPP \cite{stoyan_stochastic_2013}.
\begin{align}
\label{eqn_appendix_c_2}
&\int_{\mathbb{R}^2} \mathbb{P}(\text{SINR}_E(\mathbf{e}) \geq \theta_s) d\mathbf{e} \nonumber\\
&= 2 \pi \int_0^{\infty} \mathbb{P}(\frac{P_k h_{\mathbf{b}_0, \mathbf{e}} r^{-\alpha}}{I'_B + I'_U + I'_J + N_0} \geq \theta_s) r dr \nonumber\\
&\overset{(a)}{=} 2 \pi \int_0^{\infty} \mathbb{E}\Big[\exp(\gamma_k(r) N_0) \exp(\gamma_k(r) I'_B)\nonumber\\ &\qquad \qquad\qquad \times \exp(\gamma_k(r) I'_U) \exp(\gamma_k(r) I'_J)\Big] r dr \nonumber\\
&\overset{(b)}{=} 2 \pi \int\limits_0^{\infty} \exp(\gamma_k(r) N_0) \mathcal{L}_{I'_B}(\gamma_k(r)) \mathcal{L}_{I'_U}(\gamma_k(r)) \mathcal{L}_{I'_J}(\gamma_k(r)) r dr,
\end{align}
where $\gamma_k(r) = \frac{\theta_s r^{\alpha}}{P_k}$.
$(a)$ follows that $h \sim \exp(1)$, and $(b)$ is obtained by ignoring the correlations among $I'_{B}$, $I'_{U}$ and $I'_J$.

The point processes $\{\Phi_t\}$ and $\Phi_J$ are mutually independent.
When conditioned on $D_{\mathbf{b}_0, \mathbf{e}_i} = r$ and $\mathbf{b}_0 \in \Phi_k$, the point process of interfering BSs (i.e. $\Phi_t \setminus \mathbf{b}_0$) is the reduced Palm distribution of the PPP.
According to Slivnyak-Mecke theorem \cite{stoyan_stochastic_2013}, the reduced Palm distribution of the PPP is equal to its original distribution, i.e. $\Phi_t \setminus \mathbf{b}_0 = \Phi_t$.
Then the Laplace transforms $\mathcal{L}_{I'_B}(s)$ can be derived as 
\begin{align}
\label{eqn_appendix_c_3}
\mathcal{L}_{I'_B}(s) &= \prod_{t \in \mathcal{K}} \mathbb{E}_{\Phi_t}\Big[\prod_{\mathbf{b}_i \in \Phi_t} \mathcal{L}_h(s P_t D_{\mathbf{b}_i, \mathbf{e}}^{-\alpha})\Big] \nonumber\\
&= \exp\Big(-\frac{\pi s^{2 / \alpha}}{\sinc(\frac{2}{\alpha})} \sum_{t \in \mathcal{K}} \lambda_t P_t^{2 / \alpha}\Big).
\end{align}

One can see that the locations of the scheduled users are independent of the eavesdroppers.
Hence the point process $\Phi_U^s$ is a homogeneous PPP instead of the inhomogeneous one in the user connection probability.
According to the scheduling policy, in each slot, there is always only one user scheduled in each BS. 
Thus the intensity of $\Phi_{U, t}^s$ is $\lambda_t$, and $\lambda_{U}^{s'} = \sum_{t \in \mathcal{K}} \lambda_t$.
The Laplace transform $\mathcal{L}_{I'_U}(s)$ is 
\begin{align}
\label{eqn_appendix_c_4}
\mathcal{L}_{I'_U}(s) &= \mathbb{E}_{I'_U} \Big[\prod_{\mathbf{u}_i \in \Phi_U^s} \mathcal{L}_{h}(s P_U D_{\mathbf{u}_i, \mathbf{e}}^{-\alpha})\Big] \nonumber\\
& = \exp\Big(-\frac{\pi \lambda_U^{s'} (s P_U)^{2 / \alpha}}{\sinc(\frac{2}{\alpha})}\Big).
\end{align}

\begin{figure}[t]
	\centering
	\includegraphics[width=85mm]{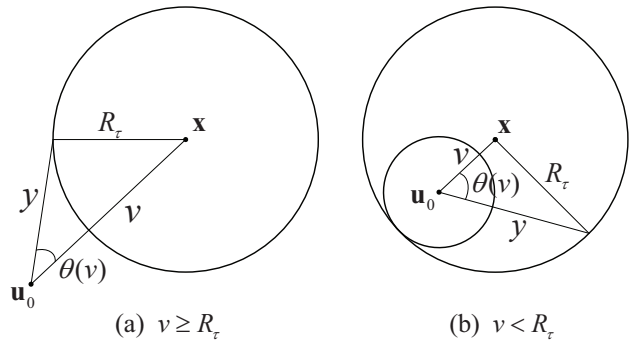}
	\vspace{-0.1cm}
	\caption{Illustration of the closest hole to the eavesdropper.}
	\label{fig_Closest_Hole}
	\vspace{-0.3cm}
\end{figure}

The locations of active jammers forms a standard PHP.
We adopt the approach proposed in \cite{yazdanshenasan_poisson_2016} and only consider the closest hole to the eavesdropper $\mathbf{e}$.
Then the Laplace transform of the interference suffered by the eavesdropper is 
\begin{align}
\label{eqn_appendix_c_5}
&\mathcal{L}_{I'_J}(s) = \mathbb{E}_{\mathbf{x}}\Big[\mathbb{E}_{\Phi_J}\Big[\prod_{\mathbf{z}_i \in \Phi_J \setminus \mathbf{B}(\mathbf{x}, R_{\tau})}\mathcal{L}_h(s P_J D_{\mathbf{z}_i, \mathbf{e}}^{-\alpha})\Big]\Big] \nonumber\\
&= \exp\Big(-\frac{\pi \lambda_J (s P_J)^{2 / \alpha}}{\sinc(\frac{2}{\alpha})}\Big) \nonumber\\
&\quad \times \mathbb{E}_{\mathbf{x}}\Big[\exp(\lambda_J \int\limits_{\mathbf{B}(\mathbf{x}, R_{\tau})} (1 - \mathcal{L}_h(s P_J y^{-\alpha})) dy)\Big] \nonumber\\
&= \exp\Big(-\frac{\pi \lambda_J (s P_J)^{2 / \alpha}}{\sinc(\frac{2}{\alpha})}\Big) \int_{0}^{\infty} G(v, s)  g(v) dv,
\end{align}
where 
\begin{align}
&G(v, s) = 
\begin{cases}
\exp\big( \int_{v - R_{\tau}}^{v + R_{\tau}} \frac{2 y\lambda_J'(y)}{1 + \frac{y^{\alpha}}{s P_J}} dy \big), v > R_{\tau},\\
\exp(\pi \lambda_J (R_{\tau} - v)^2 {}_2F_1(1, \frac{2}{\alpha}; \frac{\alpha + 2}{\alpha}; \frac{- (R_{\tau} - v)^{\alpha}}{s P_J})) \\
\quad \times \exp\big( \int_{R_{\tau} - v}^{R_{\tau} + v} \frac{2 y \lambda_J'(y)}{1 + \frac{y^{\alpha}}{s P_J}} dy \big), v \leq R_{\tau},
\end{cases}\\
& \lambda_J'(y) = \lambda_J\arccos(\frac{y^2 + v^2 - R_{\tau}^2}{2 y v}).
\end{align}
$\mathbf{x}$ is the location of the center of the closest hole (i.e. the closest scheduled user).
The first term in \eqref{eqn_appendix_c_5} is derived following a similar process to the proof of Lemma \ref{lemma_Laplace_user2user_simplified}.
The second term follows from the cosine-law: $y^2 + v^2 -2 y v \cos\theta(y) = R_{\tau}^2$ (as shown in Fig. \ref{fig_Closest_Hole}) and some geometry derivation.
Note that the centers of holes are the scheduled users whose locations are independent of $\Phi_E$.
Thus, the PDF of the distance $v = \|\mathbf{x}\|$ is given by $g(v) =2 \pi \lambda_U^{s'} \exp(-\pi \lambda_U^{s'} v^2)$.
Plugging \eqref{eqn_appendix_c_2}, \eqref{eqn_appendix_c_3}, \eqref{eqn_appendix_c_4} and \eqref{eqn_appendix_c_5} into \eqref{eqn_appendix_c_1}, Theorem \ref{theorem_secrecy_probability} is obtained.

Further discussion about the approach that only considering the closest hole of PHP could be founded in \cite{yazdanshenasan_poisson_2016}.

\bibliographystyle{IEEEtran}
\bibliography{citation}

\end{document}